\documentclass[%
 reprint,
 amsmath,amssymb,
 aps,
pra,
]{revtex4-2}

\usepackage{graphicx}
\usepackage{dcolumn}
\usepackage{bm}
\usepackage{hyperref}
\usepackage{comment}
\usepackage{xcolor}
\usepackage{upgreek}
\usepackage[utf8]{inputenc}
\usepackage[normalem]{ulem}

\begin{document}

\title{Phase diagram and universal scaling regimes of two-dimensional exciton--polariton Bose--Einstein condensates}%

\author{Félix Helluin$^1$, Daniela Pinto-Dias$^2$, Quentin Fontaine$^2$, Sylvain Ravets$^2$, Jacqueline Bloch$^2$, Anna Minguzzi$^1$, Léonie Canet$^{1}$}%
\affiliation{$^1$ Université Grenoble Alpes, CNRS, LPMMC, 38000 Grenoble France, $^2$ Université Paris-Saclay, CNRS, Centre de Nanosciences et de Nanotechnologies (C2N), 91120 Palaiseau, France}


\begin{abstract}
Many systems, classical or quantum, closed or open, exhibit universal statistical properties.  Exciton-polariton condensates, being intrinsically driven-dissipative, offer a promising platform for observing non-equilibrium universal features. By conducting extensive numerical simulations of an incoherently pumped and interacting condensate coupled to an exciton reservoir  we show that the effective nonlinearity of the condensate phase dynamics can be finely adjusted across a broad range, by varying the exciton-polariton interaction strength, allowing one to probe three main universal regimes with parameters accessible in current experiments: the weakly nonlinear Edwards-Wilkinson (EW) regime, where the phase fluctuations dominate, but the phase profile does not become rough,  the strongly non-linear Kardar-Parisi-Zhang regime, where the condensate phase fluctuations grow in a superdiffusive manner leading to roughening of the phase, and  a vortex-dominated phase
emerging at stronger interactions, where both density and phase dynamics play significant roles. Our results provide a unified picture of the phase diagram of 2d exciton-polariton condensates under incoherent pumping, and shed  light on recent experimental and numerical observations.

\end{abstract}

\maketitle

\section{\label{sec:Intro} Introduction}

Universality and scale invariance are powerful concepts of statistical physics that allow  one to classify the behaviors of physical systems in broad families independently of their microscopical details. Initially developed for systems at equilibrium, these concepts have been extended with success to the realm of out-of-equilibrium systems, where additional critical exponents arise to characterize the universal  dynamics,  and also new phenomena occur, such as self-organized criticality, where the dynamics itself drives the system to criticality without fine-tuning any external parameters.

Several universality classes have been identified in non-equilibrium critical phenomena.  Among the earliest is the directed percolation universality class in the context of reaction-diffusion processes \cite{Hinrichsen2000}.  Another celebrated universality class in non-equilibrium systems is the Kardar--Parisi--Zhang (KPZ) one, originally evidenced in randomly stirred fluids \cite{Forster76} and  in the kinetic roughening of  growing interfaces \cite{KPZ1986}. The corresponding dynamics  was successively modeled by   a stochastic continuous non-linear equation proposed by Kardar, Parisi, and Zhang \cite{KPZ1986}. KPZ processes display anomalous critical exponents and non-Gaussian distributions   of the fluctuations \cite{takeuchi2018appetizer, HHTakeuchi2015}.

{\color{black} In the  context of stochastic interface growth, another important universality class is the Edwards-Wilkinson (EW) one \cite{edwards1982surface}. It corresponds to a linear equilibrium dynamics,
 leading to non-anomalous exponents and a Gaussian distribution of fluctuations.
 However, it also emerges in  non-equilibrium dynamics such as the KPZ equation, the Kuramoto-Sivashinsky equation \cite{Kuramoto1978,Sivashinsky1977} or the deterministic complex Ginzburg-Landau equation \cite{Aranson2002} at early times, or when the system has a finite size \cite{Sneppen1992}.}

Far beyond its initial scope of application, the scale-invariant properties defining the KPZ universality class are nowadays observed in a wide variety of physical non-equlibrium phenomena, ranging from biophysics -- cancer cells proliferation \cite{KPZ_cancer_cells}, to quantum physics: integrable spin chains \cite{Znidaric_KPZ, KPZ_XXZ_Nature}, density fluctuations of an Anderson localized wavepacket  \cite{KPZ_anderson}, as well as in phase dynamics of exciton-polariton Bose-Einstein condensates  (BEC) \cite{altmanPRX2015KPZmapping}. Note that the EW universality class was also shown to emerge in such condensates \cite{Gladilin_2015_EW_to_KPZ, Mei_Wouters_scaling_2Dneq_EPBEC_2021, Helluin2024}

 Exciton-polariton  (EP) condensates are formed in optical microcavities by the strong light-matter coupling between cavity photons and excitons (electron-hole bound pairs) which, under strong enough laser driving, results in the macroscopic occupation of a single quantum state \cite{carusottoQFL}.   Exciton-polariton condensates are  in a non-equilibrium steady-state originating from the balance of pump and cavity losses. Given their driven-dissipative character, EP condensates differ radically from their equilibrium counterparts, and provide a new experimental opportunity to explore non-equilibrium universality classes.  
Remarkably, their universal features are 
directly observable from the condensate correlation functions, routinely measured in experiments. 

In the one-dimensional (1d) geometry, exciton-polaritons have been identified to belong to the KPZ universality class \cite{siebererdiehl2023universality,sieberer2016keldyshmapKPZ}
through the scaling properties of the condensate two-point correlation function in space and time \cite{Gladilin_2015_EW_to_KPZ, squizzato2018KPZsubclasses}, as well as through the identification of the KPZ universality subclasses  in the distribution of phase fluctuations \cite{squizzato2018KPZsubclasses, deligiannis2021KPZsubclasses}.
 KPZ universal features were experimentally observed in an incoherently pumped condensate realized on a 1d lattice of micropillars \cite{FontaineNature2022}. 
At variance with  a classical interface, the KPZ regime in exciton-polaritons can be hindered by the compact nature of the phase, whose $2\pi$-periodicity renders possible the formation of topological phase defects, such as space-time vortices or solitons, leading to a rich phase diagram \cite{Diehlspacetimevortex, vercesi2023phase}.

In two-dimensional (2d) systems, even more scenarii for EP universal behavior have been put forward.
As specific of the 2d geometry, a  phase transition analogous to the Berezinskii--Kosterlitz--Thouless (BKT) one \cite{Berezinsky_1970, Kosterlitz_Thouless_1972} may arise, between a quasi-ordered phase of bounded vortex-antivortex pairs, and a disordered phase of freely moving vortices. These phases are characterized respectively by algebraic and exponential decays of the condensate spatio-temporal correlation function.  The algebraic decay  was reported in experiments in quasi-equilibrium EP polaritons  \cite{Caputo2018_BKT_transition_incoherent_EPBEC}. In non-equilibrium conditions, the binding-unbinding transition of vortex-antivortex pairs has been studied theoretically building a non-linear electrodynamics analogy between vortices and charges \cite{Altman_cKPZ_duality, Altman_2016_EM_duality}. This has led to an estimate of the lengthscale $L_v$ separating attractive from repulsive interactions between vortex-antivortex pairs, as a function of the microscopic parameters \cite{Altman_2016_EM_duality}. This behavior was confirmed numerically \cite{Gladilin_Wouters_vortex_motion_2017, Gladilin_multivortex_state_2019}, and microscopic parameters such as the driving intensity are known to play a crucial role in this transition \cite{Caputo2018_BKT_transition_incoherent_EPBEC, Comaron_Noneq_BKT_2021, Marzena_2021_first_order}. Numerical studies of EP BECs close to the condensation threshold also report power-law behavior of correlations with an exponent different from the equilibrium one \cite{Comaron_Noneq_BKT_2021}.

A KPZ regime may also be found in 2d exciton-polaritons, although the conditions for achieving it are not trivial. In isotropic homogeneous incoherently pumped EP BECs, the condensate tends to be dominated by vortex proliferation which destroys the KPZ phase \cite{Altman_PRX2015_2D_superfluidity_anisotropy,Zamora_PRX2017_Tuning_across_unversalities}. However, numerical simulations have displayed the emergence of a KPZ regime in idealized conditions, {\it i.e.} in the low noise limit and in the absence of polariton-polariton interactions \cite{Mei_Wouters_scaling_2Dneq_EPBEC_2021}, or in the presence of a lattice, which introduces a cutoff length to vortices with smaller cores thereby taming the proliferation of both spatial and space-time vortices \cite{deligiannis2022KPZ2D}.
  The KPZ regime has also been predicted to occur in anisotropic condensates \cite{Altman_PRX2015_2D_superfluidity_anisotropy, Sieberer_PRL2018_defects_anisotropic_driven-open_sys, Marzena_cKPZ_2020, Ferrier_Marzena_2022}. Other pumping schemes such as coherent driving \cite{Zamora_PRX2017_Tuning_across_unversalities} or the optical parametric oscillator regime (OPO) \cite{Dagvadorj_PRX2015_BKT_OPO} also display  KPZ features.

  The goal of this work is to conciliate the above scenarii for 2d exciton-polaritons and propose a unified view of the possible universal regimes under experimentally relevant conditions.  Specifically, we consider 2d incoherently driven polariton condensates, realized on an isotropic discrete lattice and subject to interactions, which we study using extensive numerical simulations. The inclusion of this last effect in the model is not anodyne, since it gives rise to effective attractive interactions between polaritons that lead to instabilities and reduced coherence \cite{baboux2018}. In order to circumvent this effect, polariton condensates can be formed in a negative mass band resulting from the coupling of optical micropillars \cite{Schneider_2016_review_polariton_trapping}, which stabilizes the condensate and allows for large spatial coherence \cite{baboux2018}.  Indeed, this setting allowed for the observation of KPZ universality in the 1d system \cite{FontaineNature2022}. Similarly, our numerical simulations are performed setting the polaritons effective mass to a negative value.
  
We identify as tuning parameter the reservoir-polariton interaction energy, and show that when increasing its value three different regimes are achieved: the Edwards-Wilkinson, Kardar-Parisi-Zhang
and vortex phases. We then provide a detailed characterization of these three regimes. In particular, our analysis highlitghts the possibility of  experimental observation of the EW regime, characterized by power-law decay of space and time correlations,  for which we predict specific ratios of spatial and temporal power-law exponents, which are different from the equilibrium BKT case. The EW regime  is the precursor of the KPZ phase, as it emerges at small non-linearities  on time and length scales where KPZ fluctuations do not yet build up \cite{Nattermann_PRA1992_solutionEW}. Our analysis hence suggests a path towards the experimental observation of KPZ phase. Finally, we comparatively show the properties of the  vortex phase highlighting the important differences with respect to the other two phases.

\medbreak
The paper is organized as follows. The model studied, which is the generalized Gross-Pitaevskii equation (gGPE) coupled to a rate equation for the excitonic reservoir, is introduced in Sec.~\ref{sec:Model}. We then review the different universal regimes in Sec.~\ref{sec:Universal_regimes}.  Our results are presented in Sec.~\ref{sec:Results}, where we identify, and characterize in details the three distinct universal regimes, through  the first-order correlation function and its dependence on the system size, phase statistics and momentum distribution.

\section{\label{sec:Model}Model for the condensate dynamics}

Under incoherent pumping, the dynamics of a 2d weakly interacting exciton-polariton condensate can be modeled by a generalized Gross-Pitaevskii equation (gGPE) for the condensate wavefunction $\psi$ \cite{wouterscarusottoBogoBEC, wouters2009stochasticGPE}, Eq.~\eqref{eq:gGPE}. The incoherent laser driving fills high energy bands, and the condensate is indirectly populated through an excitonic reservoir of density $n_R$ \cite{wouterscarusottoBogoBEC, nature1stPolaritonBEC, baboux2018}. This process is modeled by coupling the gGPE describing the condensate dynamics to a rate equation for $n_R$ Eq.~\eqref{eq:xreservoir}, where the excitonic reservoir is incoherently driven at pumping rate $P$, scatters with polaritons with amplitude $R$, and dissipates through other decay channels at rate $\gamma_R$. The complete set of dynamical equations reads
\begin{eqnarray}
i\hbar\partial_t\psi & = & \Bigg[ \mathcal{F}^{-1}\left[\epsilon_{\hat{\boldsymbol{k}}} - \dfrac{i\hbar}{2}\gamma_{\ell}(\hat{\boldsymbol{k}}) \right] + \dfrac{i\hbar R}{2}n_R \nonumber\\
    & & + \hbar g|\psi|^2 + 2\hbar g_Rn_R  \Bigg]\psi + \hbar\xi \label{eq:gGPE} \\
\partial_tn_R & = & P - \left(\gamma_R + R|\psi|^2 \right)n_R \label{eq:xreservoir}\,,
\end{eqnarray}
where $\mathcal{F}^{-1}$ denotes the inverse Fourier transform, with $\epsilon_{\hat{\boldsymbol{k}}}$ the  dispersion relation on the lower-polariton (LP) branch. Two-dimensional polariton condensates can be realized on honeycomb lattices \cite{baboux2018}, where the LP branch takes on the characteristic shape of the graphene dispersion relation. Populating the condensate at the $\Gamma$ point of the Brillouin zone around $\hat{\boldsymbol{k}}=\boldsymbol{0}$, in the lowest energy band with anti-bonding character, the condensate dispersion relation is well approximated by the parabola $\epsilon_{\hat{\boldsymbol{k}}}=\hbar^2\boldsymbol{k}^2/(2m)$ where $m$ is the (negative) polaritonic effective mass. It can be measured from the full imaging of the LP branch before the condensation threshold. Similarly, the momentum-dependent polariton loss rate $\gamma_{\ell}(\hat{\boldsymbol{k}})$ is written within the quadratic approximation $\gamma_{\ell}(\hat{\boldsymbol{k}})=\gamma_0+\gamma_2\boldsymbol{k}^2$, where the parameters $\gamma_{0,2}$ are accessed from the condensate homogeneous spectral linewidth \cite{FontaineNature2022}. The interaction strength between polaritons is denoted by $g$, while $g_R$ represents the interaction strength between the exitonic reservoir and polaritons. The complex noise $\xi$ is Gaussian with zero mean $\langle \xi \rangle=0$ and a covariance $\langle \xi(\boldsymbol{r},t)\xi^*(\boldsymbol{r'},t') \rangle=2\sigma\delta(\boldsymbol{r}-\boldsymbol{r'})\delta(t-t')$. The amplitude $\sigma$ originates from repeated gains from the reservoir and dissipation in the environment and reads $\sigma = ({Rn_R + \gamma_0})/{4}$ \cite{carusottoQFL, wouters2009stochasticGPE}.

When the external pump $P$ is spatially homogeneous, exciton-polaritons undergo a condensation transition when $P\geq P_{\rm th} = \gamma_0\gamma_R/R$. Above this threshold, the condensate density $|\psi|^2$ is non-zero and is given at the mean-field level by $n_0=\frac{\gamma_R}{R}(p-1)$ where $p=P/P_{\rm th}$. Similarly, the mean-field excitonic density reads $n_{R0}=\gamma_0/R$.

\section{\label{sec:Universal_regimes}Universal regimes}

\subsection{\label{sec:Univ_mappingKPZ}KPZ effective dynamics}

It is now well established that the dynamics of the phase of the driven-dissipative condensate maps under some conditions onto the KPZ equation \cite{altmanPRX2015KPZmapping,Gladilin_2015_EW_to_KPZ}.
This is shown by writing the condensate wavefunction $\psi$  in density phase representation $\psi(\boldsymbol{r}, t) = \sqrt{n(\boldsymbol{r}, t)}e^{i\theta(\boldsymbol{r}, t)-i\Omega_0}$ with $\Omega_0=gn_0+2g_Rn_{R0}$ and assuming scale separation between the spatio-temporal evolution of the density and the one of the phase \cite{FontaineNature2022}. Specifically, one assumes the condensate density to be stationary and smooth enough, so that $\partial_t n \approx 0$ and $|\boldsymbol{\nabla} n| \approx 0$. Further considering that the exciton reservoir density evolves adiabatically with respect to the condensate, one finds that the effective  dynamics of the phase is governed by the KPZ equation \cite{KPZ1986}
\begin{equation}
\partial_t\theta = \nu\nabla^2\theta + \dfrac{\lambda}{2}\left( \nabla \theta \right)^2 + \sqrt{D}\eta \, ,
\label{eq:KPZmapping}
\end{equation}
where $\eta$ is a real Gaussian noise of zero mean $\langle \eta \rangle=0$ and  delta-correlated in space and time $\langle \eta(\boldsymbol{r},t)\eta(\boldsymbol{r'},t') \rangle = 2\delta(\boldsymbol{r}-\boldsymbol{r'})\delta(t-t')$. The KPZ effective parameters are given in terms  of the microscopic parameters by
\begin{align}
 \nu &= \frac{\gamma_2}{2} + \frac{\hbar g_{e}}{2mg_{i}} \label{eq:paramKPZ_nu} \\
 \lambda &= -\frac{\hbar}{m} + \gamma_2 \frac{g_{e}}{g_{i}}\label{eq:paramKPZ_lambda} \\
 D &= \frac{\sigma}{2n_0}\left[ 1+ \left(\frac{g_{e}}{g_{i}}\right)^2 \right], \label{eq:paramKPZ_D}
\end{align}
where the effective polariton-polariton interaction strength under the adiabatic approximation is denoted $g_{e}=g - 4\frac{g_Rg_{i}}{R}$ and with the losses $g_{i}=\gamma_0^2/2P$. In addition, the KPZ equation \eqref{eq:KPZmapping} can be rescaled to a unique nonlinearity parameter $g_{\rm KPZ} = \lambda^2 D/\nu^3$.

The above mapping establishes a connection between the effective  dynamics of a phase $\theta(\boldsymbol{r},t)$ and that of a classical interface of height $h(\boldsymbol{r},t)$ stochastically growing in the normal direction to the two-dimensional plane $\boldsymbol{r}$. This interface roughens and becomes scale-invariant,  characterized by a dynamical exponent $z$, a roughness exponent $\alpha$, a growth exponent $\beta=\alpha/z$, related in all dimensions by the exact identity $\alpha + z = 2$, stemming from the invariance of the KPZ equation \eqref{eq:KPZmapping} under an infinitesimal tilt of the interface \cite{MHKZ1989}. In 1d, the values of the KPZ critical exponents are exactly known $\alpha=3/2$, $z=1/2$, $\beta=1/3$. In 2d, no exact result is available but they have been determined through extensive numerical simulations of systems belonging to the KPZ class, giving the estimates $\alpha \approx 0.39$, $z\approx1.61$, $\beta\approx0.24$ \cite{Odor_directed_dmer_2DKPZ, Odor_Large_scale_simu_2DKPZ, Pagnani_exponent_2DKPZ, PERLSMAN_DP_2DKPZ, Kim_Moore_1991, Kim_Kosterlitz_1991, Tang_PRA1992_hybercube_stacking_EW-KPZ_crossover, Oliveira_2DKPZ_Lexpands, Oliveira_PRE2022_2DKP_d15}.

The interface kinetic roughening is further characterized by its two-point correlation function $C_{hh}(r,t) = \langle \left[ h(\boldsymbol{r},t)-h(\boldsymbol{0},0) \right]^2\rangle$, known to take the scaling form
\begin{eqnarray}
C_{hh}(r,t)= C_0 t^{2\beta} F\left(y_0\dfrac{r}{t^{1/z}}\right),
\label{eq:scaling_form_KPZ}
\end{eqnarray}
where $C_0$, $y_0$ are non-universal constants, and $F(y)$ is the KPZ universal scaling function, calculated in 2d using functional renormalization group \cite{Canet2010}. This function has the  following asymptotics $F(y) \underset{y\to0}{\to} F_0$ finite and  $F(y) \underset{y\to+\infty}{\sim} F_{\infty}|y|^{2\alpha}$ 
such that the correlation function behaves as a power-law  both at  coinciding space points $C_{hh}(0,t)\sim t^{2\beta}$ and equal times $C_{hh}(r,0)\sim |r|^{2\alpha}$.

Beyond its correlation properties, the universal features of the 2d growing interface can be classified to a greater extent from the distribution of its fluctuations, which indicates the existence of universality subclasses \cite{HH_PRL2012_2D_universal_distrib, Halpin-Healy_PRE2013_Pearson_distribKPZ, Oliveira_subclasses2DKPZ_2013, Carrasco_NJP_2014_2Dsubclasses_Airy}. 
These subclasses are similar to the ones demonstrated for 1d interfaces in Refs.~\cite{corwin2012KPZ, HHTakeuchi2015, takeuchi2018appetizer}. They
correspond to different global geometries of the growth:
 flat, curved or stationary. They are associated with different probability distributions although sharing the same set of critical exponents. In contrast with the 1d case, where the forms of these distributions are known exactly, being respectively  the Tracy-Widom GOE, Tracy-Widom GUE, and Baik-Rains for the three subclasses \cite{corwin2012KPZ, HHTakeuchi2015, takeuchi2018appetizer}, there is no analytical form for the 2d distributions.

\subsection{\label{sec:Univ_EW}Weakly nonlinear EW regime}

The Edwards-Wilkinson equation corresponds to the linear limit of the KPZ equation \eqref{eq:KPZmapping}, {\it i.e.} setting $\lambda = 0$ \cite{edwards1982surface}, from which the EW universality class is defined. {\color{black} The EW equation satisfies the classical limit of the fluctuation-dissipation theorem, and it describes thermal-like fluctuations around a stationary interface. }Owing from its linearity, the properties of this equation can easily be worked out, and one can obtain in generic spatial dimension $d$ the critical exponents $\alpha_{\rm EW} = \frac{2-d}{2}$, $\beta_{\rm EW} = \frac{2-d}{4}$, $z_{\rm EW}=2$ which corresponds to a diffusive behavior,  as well as its correlation functions \cite{Nattermann_PRA1992_solutionEW}.

In both one and two dimensions, the EW fixed point of Eq.~\eqref{eq:KPZmapping} is IR unstable when the nonlinearity $g_{\rm KPZ}$ is non-zero, and the renormalization group flow eventually reaches the KPZ fixed point. This means that the 2d interface always roughens. As a consequence,  when $\lambda\ll 1$, {\color{black}and for a smooth initial condition,} the scaling behavior of $C_{hh}(r,t)$ can be compatible with EW properties at small scales and at the early stages of the growth, but displays KPZ properties beyond some spatiotemporal scale which depends on $1/g_{\rm KPZ}$ \cite{Nattermann_PRA1992_solutionEW}. This scenario was confirmed from numerical simulations of lattice models of interface growth, from the gradual change of the exponents with the nonlinearity parameter \cite{Tang_PRA1992_hybercube_stacking_EW-KPZ_crossover}. Finite-size effects have also been investigated for interfaces within the EW universality class, and for the crossover between the EW and the KPZ class \cite{Tang_PRA1992_hybercube_stacking_EW-KPZ_crossover, Forrest1990, PAL1999_EW2+1}.

The EW universality class is also relevant in the context of the complex Ginzburg-Landau equation (CGLE). While this  deterministic equation is known to host a wide variety of patterned structures such as spirals, vortex glasses, as well as spatiotemporal chaos \cite{Aranson_RMP2002_world_CGLE,Chate_PhysicaA1996_phase_diagram_2DCGLE}, it also exhibits a regime,   called phase turbulence, where the phase dynamics can be mapped into the Kuramoto-Sivashinsky equation~\cite{Kuramoto1978,Sivashinsky1977}. This deterministic equation has unstable modes which leads to chaos, such that its  effective large-distance properties are known to belong to the KPZ universality class  \cite{Hyman1986,Cuerno1995}.
 However, in practice,  huge system sizes and simulation times are required to actually observe the KPZ behavior, and otherwise the physical properties of the system display the EW universal features  \cite{Chate_PhysicaD1996_phase_turbulence_deterministic_2DCGLE_finite_size,Pandit_PhysicaD1996_conjeture_PT_CGLE_finite_size,Roy2020,Vercesi2024}.

\subsection{\label{sec:Univ_defects}Defects-dominated regime}

A major difference between the usual KPZ equation \cite{KPZ1986} and the one obtained in Eq.~\eqref{eq:KPZmapping} lies in the compactness of the phase $\theta(\boldsymbol{r},t)$, which is defined on the support $\left[-\pi,\pi \right[$, whereas the interface $h(\boldsymbol{r},t)$ grows unbounded in $\mathbb{R}$. This specific feature has deep consequences on the growth of the phase  $\theta(\boldsymbol{r},t)$, as it renders possible the formation and proliferation of topological defects. Their nucleation has been studied building a dual electrodynamics theory for the compact version of the KPZ equation \cite{Altman_cKPZ_duality}, and it was shown from a perturbative treatment in the nonlinearity $\lambda$ that vortex-antivortex pairs should undergo a binding-unbinding mechanism, separating an algebraic quasi-ordered phase from a disordered one \cite{Altman_2016_EM_duality, Sieberer_PRL2018_defects_anisotropic_driven-open_sys}. This mechanism is characterized by a lengthscale $L_v$ representing the typical defect interdistance, that depends on the effective parameters $\nu$, $\lambda$ and $D$ of the KPZ equation. Numerical simulations of the compact KPZ equation indicate that defect nucleation is hindered for systems smaller than $L_v$ \cite{Marzena_cKPZ_2020}.

\medbreak
 The deterministic part of the gGPE \eqref{eq:gGPE} describing the condensate dynamics is well approximated by the 2d CGLE. At variance with the  vortices in the compact version of the KPZ equation, the phase defects of the CGLE are pinned to density holes, that are continuously replenished by the external drive. This causes an outward polariton flow from these holes, which in return dramatically affects the interaction between vortex pairs \cite{Gladilin_Wouters_vortex_motion_2017, Gladilin_multivortex_state_2019}, creating notably self-accelerated motion of the vortex cores \cite{Gladilin_Wouters_vortex_motion_2017, Aranson_PRL1994_self-acceleration_spiral_vortex}.

However, the stochastic part of the gGPE \eqref{eq:gGPE} is known to modify the phase diagram of the 2d CGLE, e.g destroying vortex glasses, affecting the motion of spiral vortices, and creating additional free vortices \cite{Chate_PhysicaA1996_phase_diagram_2DCGLE, Chate_PRL1998_Spiral_motion_noisy_2DCGLE}. It was also pointed out that phase defects can appear in the space-time map of the phase \cite{Diehlspacetimevortex}. These space-time vortices are noise-activated, and their density was found to be proportional to $e^{-\frac{p-1}{\sigma}}$ in 1d \cite{vercesi2023phase}. While   KPZ universal features were shown to be robust to small defect densities in both 1d and 2d condensates \cite{FontaineNature2022, vercesi2023phase, deligiannis2022KPZ2D}, they are superseded by  new properties in the vortex dominated phase.

\section{\label{sec:Results} Results}

\subsection{\label{sec:Method_Num}Numerical simulations}

We solve numerically the gGPE \eqref{eq:gGPE} with periodic boundary conditions by propagating the initial condition in time with a symmetrized split-step method \cite{agrawal2001}.  We implement the stochastic part of the dynamics
by adding to the wave function $\psi(\boldsymbol{r},t)$, at each time step and for each spatial point, a complex random number generated by a Euler-Maruyama algorithm, reproducing the statistical properties of the Gaussian noise $\xi$. We solve the reservoir dynamics \eqref{eq:xreservoir} by using a first order Euler scheme. {\color{black}The codes and data produced are available in Ref.~\cite{helluin_zenodo}.}

The parameters of the simulation are chosen to reproduce typical experimental conditions in GaAs microcavities~\cite{FontaineNature2022}: $\hbar\gamma_0=197.5{\rm\mu eV}$, $\hbar\gamma_2=1.77\times10^4{\rm\mu eV.\mu m^{2}}$, $\gamma_R=0.45\gamma_0$, $R=57.3\times10^{-3}{\rm\mu m.ps^{-1}}$, $P=2P_{\rm th}$. The polariton-polariton interaction strength $g$ is set to zero, and the interaction between the exciton reservoir and polaritons $g_R$ is chosen from the value of the  blueshift at threshold $\mu_{\rm th}$ and the steady-state reservoir density, $g_R = \frac{1}{2}\mu_{\rm th}/n_{R0}$. The blueshift at threshold $\mu_{\rm th}$ is varied from $50{\rm \mu eV}$ to $700{\rm \mu eV}$ in order to access  the different universal regimes presented in Sec.~\ref{sec:Universal_regimes}.

{\color{black}Note that, in previous works where KPZ universality was evidenced in 2D condensates \cite{Mei_Wouters_scaling_2Dneq_EPBEC_2021, deligiannis2022KPZ2D}, the condensate dynamics was studied in the adiabatic approximation, neglecting the reservoir dynamics. Moreover, in Ref. \cite{Mei_Wouters_scaling_2Dneq_EPBEC_2021}, the parameter $\gamma_2$, which plays a crucial role in the stability of the condensate, was set to zero, and the study was performed in the low noise limit. In Ref. \cite{deligiannis2022KPZ2D}, $\gamma_2\neq 0$ and the noise was nominal, but the interactions were set to zero. Here, we consider non-zero interactions, and the full non-adiabatic description. Beside being more  realistic from an experimental viewpoint, setting a non-zero interaction strength $g_R$ dramatically affects the phase diagram of positive mass polariton condensates, creating unstable modes even in the adiabatic approximation of the reservoir.}
The polariton mass is thus chosen negative $m=-4\times10^{-5}m_e$ so as to counterbalance the sign of the effective interactions ($g_e<0$) and 
stabilize the condensate \cite{baboux2018}. {\color{black} Moreover, non-adiabatic effects in the exciton reservoir can give rise to yet other unstable modes or to temporal modulation of the first order correlation function \cite{vercesi2023phase}. Including all these elements in our numerics allows us to explore a rich and diverse phase diagram, while addressing the question of the accessibility of KPZ features in the most general setting.
} The noise amplitude is approximated in the following by $\sigma \approx Rn_R/2$.

We perform the numerical simulations on a square lattice of spacing $\ell = 2.83{\rm\mu m}$ to mimic the structure of a square lattice of $N\times N$ optical micropillars, and the number of sites is $N = 64$ (unless otherwise mentioned). The integration time-step $dt$ is varied linearly with the blueshift at threshold $\mu_{\rm th}$ from $3\times10^{-2}{\rm ps}$ to $8\times10^{-3}{\rm ps}$, such that the phase dynamics remains well resolved in time. The solution of the stochastic condensate dynamics Eq.~\eqref{eq:gGPE} reaches a steady state  at a time $t_0$ which depends on the chosen microscopic  parameters. In the steady state, the spatially averaged density is stationary. In the remainder of this article, we focus on the steady-state properties of the condensate for times larger than $t_0$.

\subsection{\label{sec:Method_phase_diag} Identification of the three regimes}

In the following, we show that  the three regimes presented in Sec.~\ref{sec:Universal_regimes} are accessible by tuning the nonlinear parameter $g_{\rm KPZ} = \frac{\lambda^2 D}{\nu^3}$, whose dependence on the microscopic parameters is known from Eqs.~\eqref{eq:paramKPZ_nu}--\eqref{eq:paramKPZ_D}. This effective nonlinearity can be tuned over a broad range of values  by changing the blueshift at threshold $\mu_{\rm th}$, which controls the effective polariton interaction strength $g_{e}$. {\color{black} Experimentally, the blueshift at threshold can be modified by changing the spectral detuning between the exciton resonance and the cavity mode, thus tuning the excitonic fraction of the polariton state \cite{Schneider_2016_review_polariton_trapping, comaron_ostrovskaya_arXiv2024_algebraic_g1-r}}. It thus provides a natural candidate to explore the universal properties of the 2d condensate. {\color{black}Note that similar results could be  obtained varying the pumping rate $p$ instead. }

\begin{figure}[ht!]
\includegraphics[width=9cm]{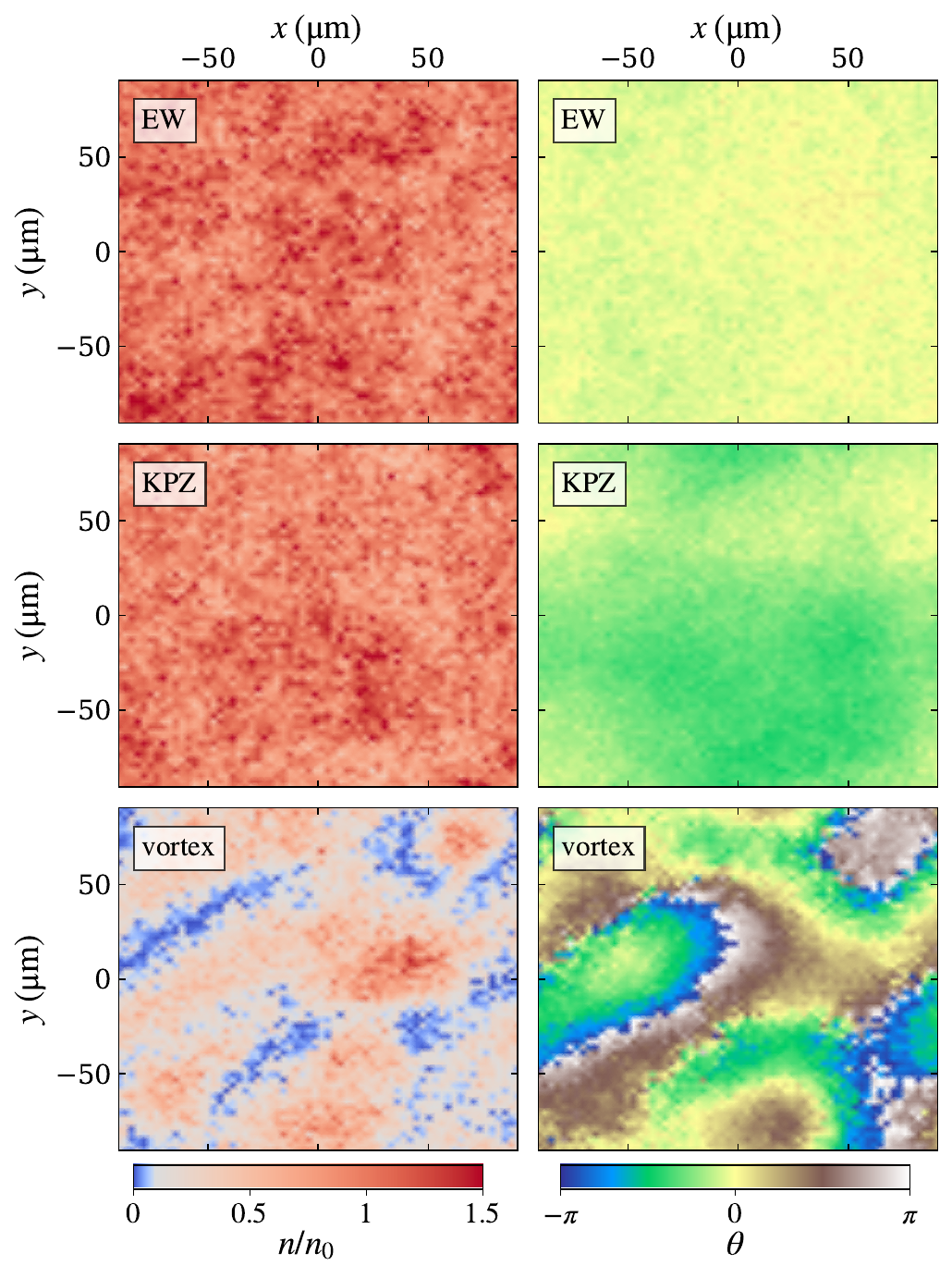}
\caption{\label{fig:2Dmaps} Typical stationary density (left column) and phase (right column) maps, for an arbitrary time instant in the steady-state, in the three different universal regimes: (EW) $\mu_{\rm th}=50\mu eV$, (KPZ) $\mu_{\rm th}=429\mu eV$, (vortex) $\mu_{\rm th}=700\mu eV$.  The condensate density $n(x,y)$ is normalized by its mean field value $n_0$.}
\end{figure}
We solve the condensate dynamics Eqs.~\eqref{eq:gGPE}--\eqref{eq:xreservoir} for many values of $\mu_{\rm th}$, and
 indeed identify three distinct regimes, which correspond to the EW, the KPZ and the vortex regimes of Sec.~\ref{sec:Universal_regimes}.
We first provide a qualitative picture, by displaying in Fig.~\ref{fig:2Dmaps} typical  density and phase maps, which are snapshots at an arbitrary instant in the steady state, for  each of these regimes.  The density (left column) appears almost identical in both the EW (upper row) and the KPZ (middle row) regimes. In the vortex phase (bottom row), the condensate density visibly spreads out toward lower values, and interacting vortices are discernible. Regarding the phase (right column), a clear difference is however noticeable between  the EW  and KPZ  regimes from the spread of the phase values. In the vortex regime, the phase explores entirely its compact support and phase defects are visible at spatial points mixing all the colors together, and coinciding with  holes in the density map, indicating a strong correlation between the density and phase fields. We have checked that the phase diagram remains similar decreasing the pumping rate down to $p\simeq1.5$, below which the proliferation of space-time defects also affects the condensate dynamics \cite{Diehlspacetimevortex, vercesi2023phase}.

\subsection{\label{sec:Method_phase_diag} Steady-state condensate density}

We compute for a wide range of interaction strength the steady state density, averaged over space and noise realizations. Simulations are performed over $10^6$ps to let the condensate relax to its steady state, and the numerical convergence is sped up by sampling each initial condition from the steady state configuration reached by the condensate for the previous value of $\mu_{\rm th}$.

\begin{figure}[ht!]
\includegraphics[width=7cm]{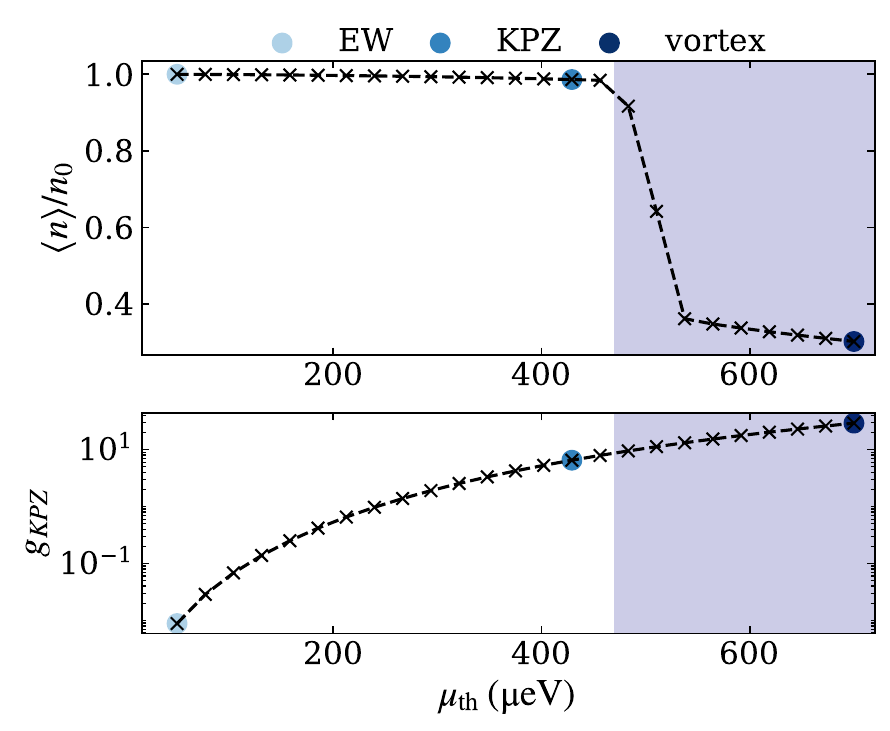}
\caption{\label{fig:transition} Condensate density averaged over space and $128$ realizations of the noise, normalized by $n_0$ (upper panel) and effective KPZ nonlinearity defined from Eq.~\eqref{eq:KPZmapping}  (lower panel) as a function of the blueshift at threshold $\mu_{\rm th}$. Light, medium, and dark blue circles indicate points of the phase diagram corresponding to the EW, KPZ, and  vortex regime respectively. }
\end{figure}

Results are shown as a function of the blueshift at threshold on the upper panel of Fig.~\ref{fig:transition}, revealing a sharp transition at $\mu_{\rm th}\approx 480{\rm\mu eV}$. This abrupt drop of density is related to  the transition from a vortex-free condensate with very few space-time vortices to a defect-dominated phase (blue-shaded area). A similar transition was reported in  Ref.~\cite{Marzena_2021_first_order} as a result of modulational instability \cite{Ostrovskaya_PRB2014_modulation_instability_dark_solitons, Ostrovskaya_PRB2015_modulation_instability_2DspinorBEC, baboux2018, vercesi2023phase}, whereas in the present case it rather seems to be related to the competition between the finite system size and the typical vortex separation scale $L_v$ \cite{Marzena_cKPZ_2020}, establishing a transition from a phase turbulent to a defect turbulent regime \cite{Chate_PhysicaD1996_phase_turbulence_deterministic_2DCGLE_finite_size}.

 The precise location of the transition point is blurred out due to finite-size effects, and to the fact that we observe extended relaxation timescales  in the presence of vortices, which are consistent with the ones reported in Ref.~\cite{Ferrier_Marzena_2022}. The slow relaxation induced by the presence of vortices  was also noted in the relaxation dynamics of the equilibrium XY model \cite{Jervis_PRL2000_relaxation_vortices_XY}. Thus, in order to confirm  the position of the defect phase indicated by the blue-shaded area, we ran the algorithm from both sides, {\it i.e} for both increasing and decreasing $\mu_{\rm th}$.

We also show on Fig.~\ref{fig:transition} (lower panel) the corresponding values of the effective KPZ non-linearity  $g_{\rm KPZ}$. EW properties are expected to be dominant in the weakly nonlinear regime reached for small $\mu_{\rm th}$, followed by KPZ properties at intermediate $\mu_{\rm th}$, before the  vortex-dominated phase sets in.
Three points representative of these three regimes are indicated on Fig.~\ref{fig:transition}. They correspond to the three values chosen for the  maps of Fig.~\ref{fig:2Dmaps}.

In the following,   we characterize the properties of these three regimes  through their first-order coherence, their momentum distribution, and their fluctuations statistics.

\subsection{\label{sec:Results_g1}First-order correlation \texorpdfstring{$g^{(1)}(r,t)$}{g1}}

Each of the regimes presented in Sec.~\ref{sec:Universal_regimes} deeply imprints the phase dynamics of the EP condensate, and therefore its coherence properties. They can be distinguished  experimentally  in the spatiotemporal decay of the condensate first-order correlation function, defined as  
 \begin{equation}
 g^{(1)}_{\psi}(\Delta r,\Delta t)=\frac{\langle \psi^*(\Delta\boldsymbol{r} + \boldsymbol{r_0},\Delta t+t_0)\psi(\boldsymbol{r_0},t_0) \rangle}{\sqrt{\langle n(\Delta\boldsymbol{r}+\boldsymbol{r_0},\Delta t + t_0) \rangle\langle n(\boldsymbol{r_0},t_0) \rangle}}\,
 \label{eq:g1_def}
 \end{equation}
where the average $\langle\cdot\rangle$ is performed over realizations of the stochastic dynamics \eqref{eq:gGPE} in the steady state,  as well as over the circle of radius $\Delta r=|\Delta \boldsymbol{r}|$ exploiting isotropy. This correlation function encompasses contributions from both the density and the phase, and approximately reads
\begin{equation}
     g^{(1)}_{\psi}(\Delta r, \Delta t) \approx g^{(1)}_{n}(\Delta r, \Delta t)\langle e^{i\left[\theta(\Delta\boldsymbol{r} + \boldsymbol{r_0}, \Delta t + t_0) - \theta(\boldsymbol{r_0}, t_0)\right]} \rangle
\end{equation} 
when density-phase correlations are negligible,  with
\begin{equation}
     g^{(1)}_n(\Delta r, \Delta t) = \dfrac{\langle \sqrt{ n(\Delta\boldsymbol{r} + \boldsymbol{r_0},\Delta t+t_0) n(\boldsymbol{r_0},t_0)}\rangle}{\sqrt{\langle n(\Delta\boldsymbol{r} + \boldsymbol{r_0},\Delta t+t_0) \rangle\langle n(\boldsymbol{r_0},t_0) \rangle}}.
\end{equation} 
This is the case  both in the EW and in the KPZ regimes.
Additionally assuming that density-density correlations are negligible $g^{(1)}_n(\Delta r, \Delta t)\approx1$ and performing a cumulant expansion  of the phase term one obtains \cite{FontaineNature2022}
 \begin{equation}
 -2\log\left[ |g^{(1)}_{\psi}(\Delta r, \Delta t)| \right] \simeq C_{\theta\theta}(\Delta r,\Delta t)
 \label{eq:cumulant}
 \end{equation}
where $C_{\theta \theta}(\Delta r, \Delta t)$ is the connected two-point correlation function of the phase defined as for the KPZ height-height correlator Eq.~\eqref{eq:scaling_form_KPZ}. In the following we adopt the short-hand notation $\mathcal{G}^{(1)}_{\psi, n}(\Delta r,\Delta t)\equiv -2\log\left[ |g^{(1)}_{\psi, n}(\Delta r, \Delta t)| \right]$.

\begin{figure}[ht!]
\includegraphics[width=8cm]{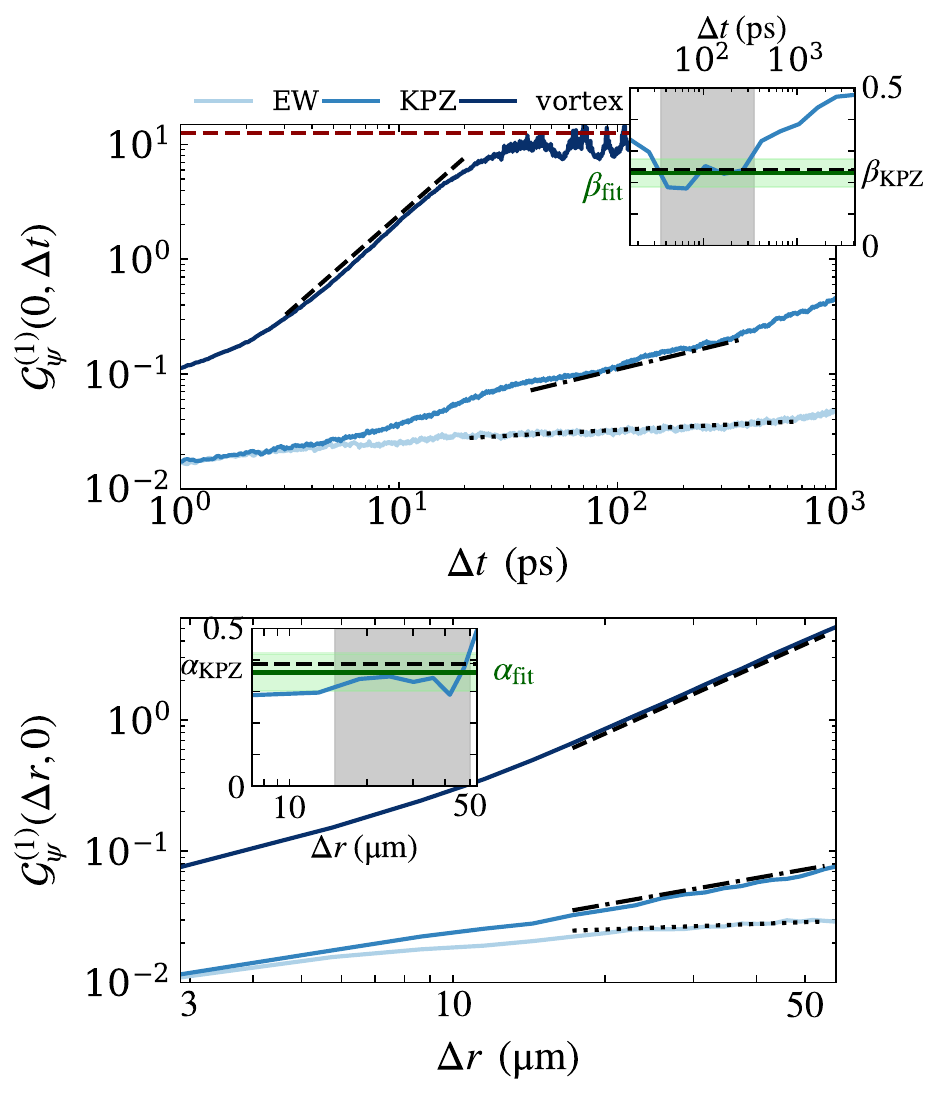}
\caption{\label{fig:g1} Equal-space (upper panel) and equal-time (lower panel) first-order correlation function corresponding to the EW, KPZ, and vortex points of Fig.~\ref{fig:transition}. A logarithmic fit is plotted in dotted line for the former, and guidelines are shown in dash-dotted and dashed line for the two latters. The red horizontal dashed line corresponds to the expected saturation value of $g^{(1)}(0,\Delta t)$ in the absence of density correlations and for $\text{Var}\left[\theta(r_0, \Delta t + t_0)-\theta(r_0, t_0)\right]=2\pi$. The local exponents, computed according to Eqs.~\eqref{eq:loc_expo_r}--\eqref{eq:loc_expo_t}, are shown for the KPZ point in the insets, where the grey shade extends, as the dash-dotted line, over $\left[35{\rm ps} , 350{\rm ps}\right]$ and $\left[15{\rm\mu m}, 50{\rm\mu m} \right]$ respectively.}
\end{figure}

It follows from the mapping \eqref{eq:KPZmapping} and \eqref{eq:cumulant} that if the phase of the condensate obeys at some spatiotemporal scales a KPZ dynamics, the coherence of the condensate endows the  scaling form 
\begin{equation}
\mathcal{G}^{(1)}(\Delta r, \Delta t)\sim \left\{\begin{array}{l l}
                                            \Delta t^{2\beta_{\rm KPZ}}, \qquad& \Delta r=0\\
                                            \Delta r^{2\alpha_{\rm KPZ}}, & \Delta t=0
                                        \end{array}\right. \,,
                                           \label{eq:KPZscaling}
\end{equation}
as suggested in early studies of the CGLE \cite{Pandit_PhysicaD1996_conjeture_PT_CGLE_finite_size}. 
As already mentioned in Sec.~\ref{sec:Intro}, this behavior has been confirmed in experimental realizations of 1d polariton condensates  \cite{FontaineNature2022}, as well as in numerical simulations of their 2d analog {\color{black} under idealized conditions} \cite{Mei_Wouters_scaling_2Dneq_EPBEC_2021, deligiannis2022KPZ2D}.

For weak but non-vanishing nonlinearity, {\color{black} although the system is out-of-equilibrium}, EW correlations are expected to persist for extended scales before KPZ features arise. While stretched exponential decays of $g^{(1)}$ are expected in 1d in the EW phase as in the KPZ phase \cite{Helluin2024}, the situation changes in 2d as a result of the vanishing of $\alpha_{\rm EW}$ and $\beta_{\rm EW}$ (Sec.~\ref{sec:Univ_EW}). This yields instead logarithmic growth as predicted from the solution of the EW equation \cite{Nattermann_PRA1992_solutionEW}
\begin{equation}
\mathcal{G}^{(1)}(\Delta r, \Delta t)\sim \left\{\begin{array}{l l}
                                            \dfrac{D}{2\pi\nu}\ln(\Delta t\Lambda^2), \qquad& \Delta r=0\\
                                            \dfrac{D}{\pi\nu}\ln(\Delta r\Lambda), & \Delta t=0
                                        \end{array}\right. \,,
                                           \label{eq:EWscaling}
\end{equation}
with $\Lambda$ a UV cutoff. This implies an algebraic decay of $g^{(1)}$ in space as $g^{(1)}(\Delta r,0)\sim \Delta r^{-a_s}$ with  exponent $a_s=D/2\pi\nu$, and in time as $g^{(1)}(0,\Delta t)\sim \Delta t^{-a_t}$ with exponent $a_t=D/4\pi\nu$. Such a behavior was predicted within Bogoliubov theory \cite{chiocchetta2013, Marzena2010_finite_size} and observed in experiments \cite{Roumpos_PNAS2012_algebraic_decay}
but however never related to EW universal features.
We highlight that the EW power-law exponents are different from the 2d 
{\color{black}closed case}, where the BKT ordered phase displays an exponent ratio $a_s/a_t=1$. Values  $a_s/a_t\simeq 1$ were reported in experiments on quasi-equilibrium polariton condensates \cite{Caputo2018_BKT_transition_incoherent_EPBEC}.
The identification of the EW regime, motivated by the mapping \eqref{eq:KPZmapping} as well as results shown on Fig.~\ref{fig:transition}, appears crucial in guiding the observation of KPZ universality in 2d polariton condensates. Moreover, this result provides a quantitative explanation for the exponent ratio estimated numerically as $a_s\simeq 2a_t$ in Ref. \cite{Comaron_Noneq_BKT_2021}, as well as for the recent experimental observation \cite{comaron_ostrovskaya_arXiv2024_algebraic_g1-r} that $a_s\propto1/n_0$.

Let us note that in Ref.~\cite{comaron_ostrovskaya_arXiv2024_algebraic_g1-r},  it was suggested that 
{\color{black}$a_s\propto 1/n_0\lambda_T^2$} in analogy with equilibrium condensates {\color{black}with $\lambda_T = \hbar/\sqrt{2\pi m k_B T}$ the thermal de-Broglie wavelength. This expression is however   ambiguous as a temperature $T$ is not defined out of thermal equilibrium. An effective temperature can however be rigorously defined, comparing the tails of the polariton condensate momentum distribution with that of a thermal closed Bose gas \cite{chiocchetta2013}, where in our notations $k_BT_{\rm eff} = \frac{2D}{\nu}\frac{\hbar^2n_0}{2m}$. The spatial EW exponent can in turn be rewritten $a_s = 1/n_0\lambda_{T_{\rm eff}}^2$, making the \color{black}{non-equilibrium} nature of the condensate perfectly explicit since the effective temperature depends on drive and dissipation, \color{black}{and the noise is not constrained by any fluctuation-dissipation type of relation}. Note that to take into account actual thermal effects,  an additional phonon bath can be included  and it is found to contribute to the noise strength $\sigma$ \cite{anna-maxime, Shelykh}. }
The general correspondence between Bogoliubov prediction for the large space and time behavior of the phase correlation and EW properties is detailed in Appendix~\ref{sec:ap_bogo}.

\begin{figure}[ht!]
\includegraphics[width=8cm]{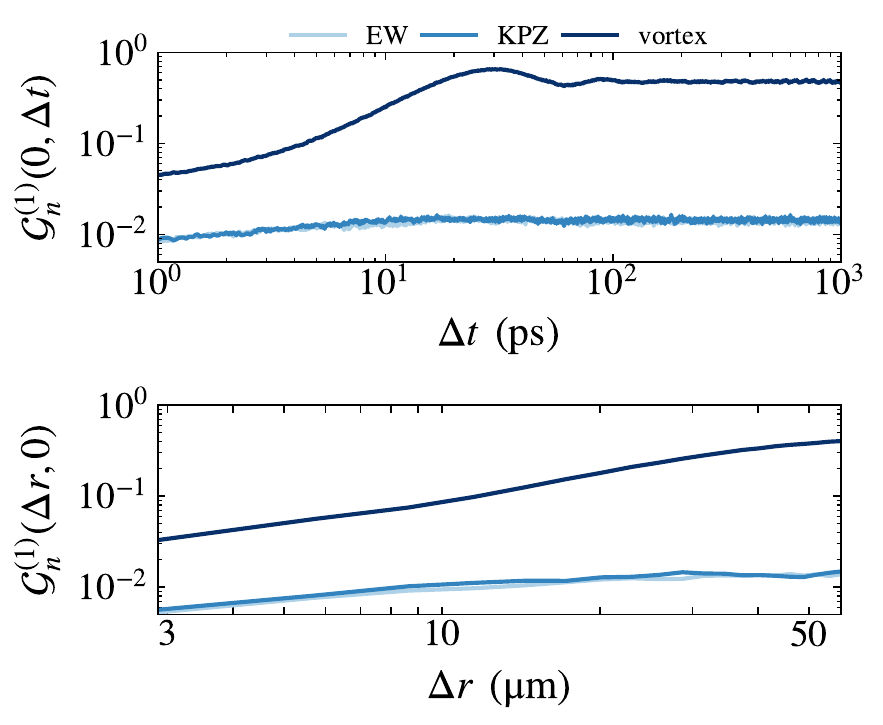}
\caption{\label{fig:g2} Equal-space (upper panel) and equal-time (lower panel) density correlation.}
\end{figure}

\medbreak
The first-order correlation function is computed in the steady state of the condensate according to Eq.~\eqref{eq:g1_def}, and averaged over 1024 realizations of the noise. The results obtained in the three regimes are shown on Fig.~\ref{fig:g1}. In the weak nonlinearity regime (light blue), algebraic decays consistent with EW behavior \eqref{eq:EWscaling} are observed, and the corresponding fits are shown as dotted line, where we find $a_s/a_t=1.8\pm 0.1$. These fits are performed in a restricted spatiotemporal window (grey-shaded area). Its lower bound corresponds  to the time where density correlations (second-order correlation function), displayed in  Fig.~\ref{fig:g2},  are negligible ($\mathcal{G}^{(1)}_n$ is flat) such that the first-order correlation function is fully determined by the phase. Its upper bound corresponds to the time  when $g^{(1)}(0, \Delta t)$ develops an exponential decay induced by the finite system size (discussed in Sec.~\ref{sec:Results_finite_size}).

At intermediate values of $\mu_{\rm th}$ (medium blue), a clear departure from EW behavior is observed. This departure is induced by phase correlations, as shown on Fig.~\ref{fig:g2}, where the decay of density correlations perfectly matches the one at the EW point, indicating that approximations leading to the effective phase dynamics \eqref{eq:KPZmapping} are valid. Reservoir effects are visible at times $\Delta t\approx 10$ to $30$ps and naturally arise when interactions are increased  or for high values of the pumping rate \cite{vercesi2023phase, deligiannis2022KPZ2D}. Proceeding as in Ref.~\cite{deligiannis2022KPZ2D}, we compute the local exponents according to
\begin{subequations}
\begin{align}
    \alpha & \approx \frac{1}{2}\dfrac{d}{d\ln(\Delta r)}\ln\left[\mathcal{G}^{(1)}(\Delta r,0)\right]  \label{eq:loc_expo_r}\\
    \beta & \approx \frac{1}{2}\dfrac{d}{d\ln(\Delta t)}\ln\left[\mathcal{G}^{(1)}(0,\Delta t)\right].\label{eq:loc_expo_t}
\end{align}
\label{eq:loc_expo}
\end{subequations}
Results are shown in the insets of Fig.~\ref{fig:g1}. The exponents are extracted within the windows of interest, {\it i.e} after density or reservoir effects and before finite size effects. We find $\beta_{\rm fit} = 0.23\pm0.04$ and $\alpha_{\rm fit} = 0.36\pm 0.06$, which are in good agreement with exponents obtained from numerical resolution of discrete models in the 2d KPZ universality class  \cite{Pagnani_exponent_2DKPZ}. These fitted exponents are used to plot the dash-dotted guidelines on Fig.~\ref{fig:g1}, in windows indicated as the grey shades over 315ps in time and 35${\rm\mu m}$ in space.

At large values of $\mu_{\rm th}$ (dark blue), the first-order correlation function decays very rapidly in both space and time due to the proliferation of phase defects, and saturates in time to $g^{(1)}\approx e^{-2\pi}$ when the compact phase completely spreads in $\left[ -\pi, \pi \right]$. Stretched exponential decays of the form of \eqref{eq:KPZscaling} are indicated by dashed guidelines, with $\alpha_{v} = 0.857\pm0.005$ and $\beta_{v}= 0.835\pm0.003$, which lie within the range of  values reported in Ref.~\cite{Caputo2018_BKT_transition_incoherent_EPBEC} while clear exponential decays ($\alpha_v=\beta_v=0.5$) were found in Ref.~\cite{Comaron_Noneq_BKT_2021}. It should also be noted that density correlations exhibit oscillations in time (upper panel Fig.~\ref{fig:g2}), which are found independent on $g_R$ but sensitive on the pumping rate $p$. This numerical observation could be of experimental relevance for the characterization of the defect regime, and calls for further theoretical investigations that go beyond the scope of this work.

\subsection{\label{sec:Results_finite_size}Finite-size effects}

\begin{figure}[ht!]
\includegraphics[width=8cm]{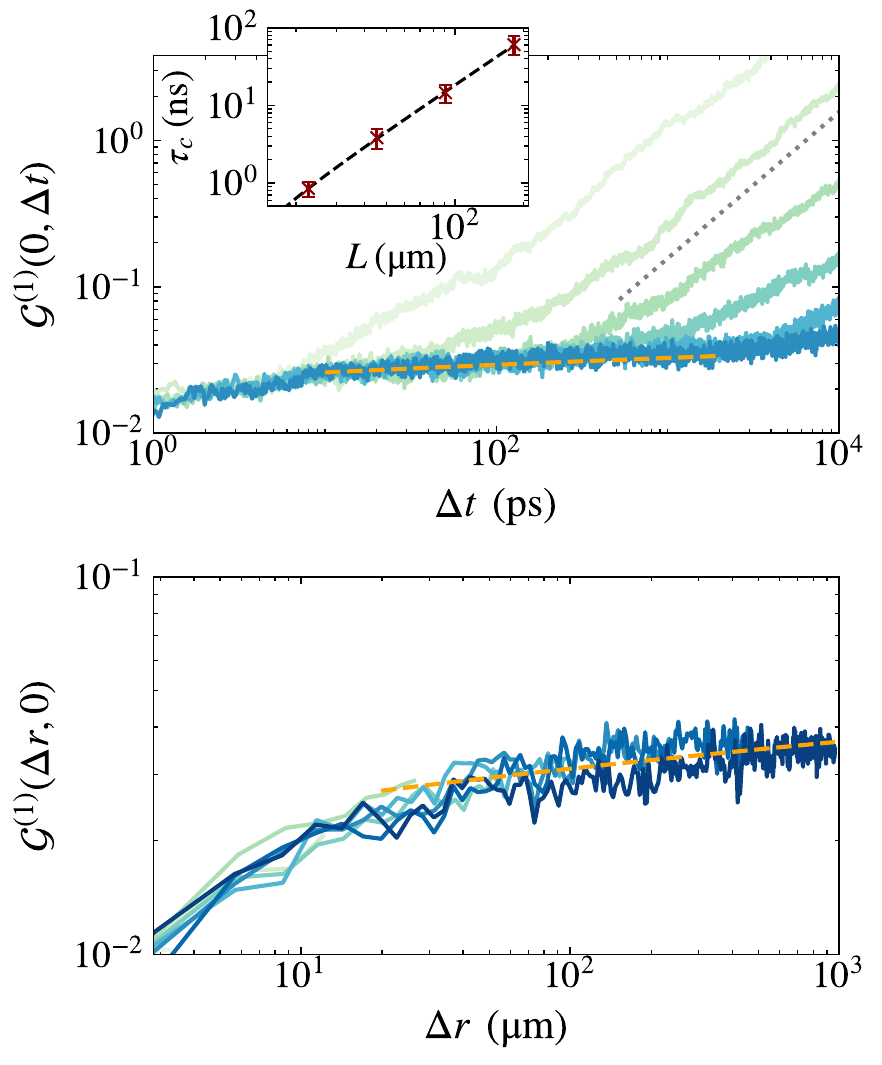}
\caption{\label{fig:finite_size_EW} First-order correlation function at the EW point of Fig.~\ref{fig:transition} for a system of $N=2^i$ linear sites, at (upper panel) equal-space with $i=\{3,\dots,8\}$, and (lower panel) equal-time with $i=\{3,\dots,10\}$. The colorscale is changing with the increasing system size from light green to dark blue. Algebraic decays are represented by the fitted orange dashed line. A grey dotted line indicates the large time exponential decay. Inset: coherence time $\uptau_c$ fitted from the large-time exponential decay of $g^{(1)}(0,\Delta t)$ as a function of the system size $L$.  }
\end{figure}

In addition to the scaling laws expected in the decay of the coherence, we show below that KPZ or EW critical exponents also imprint the exponential decay superseding the EW-KPZ regimes, through finite-size effects. We thus  now investigate this exponential decay, which can be observed  on Fig.~\ref{fig:g1} in the equal-space first-order correlation function at the latest times. For this, we compute the condensate first-order correlation function averaged over 256 realizations of the noise at the EW point of Fig.~\ref{fig:transition} increasing the system size $N\times N$ from $N=8$ to $N=256$ for temporal correlations and up to $N=1024$ for spatial correlations. Results are shown on Fig.~\ref{fig:finite_size_EW}, where one observes that the condensate coherence preserves a quasi long-range order down to millimeter lengths. As a function of time, the condensate exhibits quasi long-range order up to a crossover time, that depends on the system size, from which the decay becomes exponential. This result can be summarized in the following scaling form for the equal-space phase-phase correlator
\begin{equation}
    C_{\theta\theta}(0,t) \sim  t^{2\beta}f\left( \dfrac{t}{L^z} \right),
    \label{eq:scaling_finite_size}
\end{equation}
which was first formulated in the context of kinetically roughening interfaces \cite{Krug_AP1997_finite_size}, then suggested in the context of the CGLE \cite{Pandit_PhysicaD1996_conjeture_PT_CGLE_finite_size}, and very recently again in the large time decoherence of 1d driven-dissipative condensates \cite{amelio2024KPZlinewidth}. A derivation of Eq.~\eqref{eq:scaling_finite_size} is presented within Bogoliubov theory in both one and two dimensions in Appendix~\ref{sec:ap_bogo}. This equation represents the change in behavior of correlations between their thermodynamics limit and their saturation for finite size, after the growing interface (here the phase) has lost the memory of its initial condition \cite{Krug_AP1997_finite_size}.

In 2d and in the weakly nonlinear regime where $z=2$, one finds $f(t/L^z)\underset{t\ll L^z}{\sim}\log(t/L^2)$ which reproduces the algebraic decay of the condensate coherence in Eq.~\eqref{eq:EWscaling}. At large times $t/L^z\gg 1$, $f(y)\sim y^{1-2\beta}$, from which we obtain $g^{(1)}(0, \Delta t)\sim e^{-\Delta t/\uptau_c}$ with the coherence time $\uptau_c \sim L^{z-2\alpha}$. This means that while the condensate decoherence is constrained to an exponential decay at large times no matter $z$ and $\alpha$, the coherence time keeps the memory of the underlying universality class through its dependence on the system size. This could be of great importance to identify KPZ universal features in 2d polariton condensates given the limited extent of the temporal scaling in its one-dimensional experimental realization \cite{FontaineNature2022}, and notably in our case due to finite size effects as seen on Fig.~\ref{fig:g1}.
In the weakly nonlinear limit under consideration here we expect $\uptau_c\sim L^2$, which coincides with the Schawlow-Townes coherence time of 2d lasers \cite{Schawlow_1958, Fabre_2010}. The coherence time $\uptau_c$ is extracted from the exponential decay of Fig.~\ref{fig:finite_size_EW} and plotted in the inset. Fitting these values with a power law ansatz, we find $\uptau_c\sim L^{1.98\pm0.07}$ which is in perfect agreement with the expected  EW   scaling  exponents.

\medbreak
\begin{figure}[ht!]
\includegraphics[width=8cm]{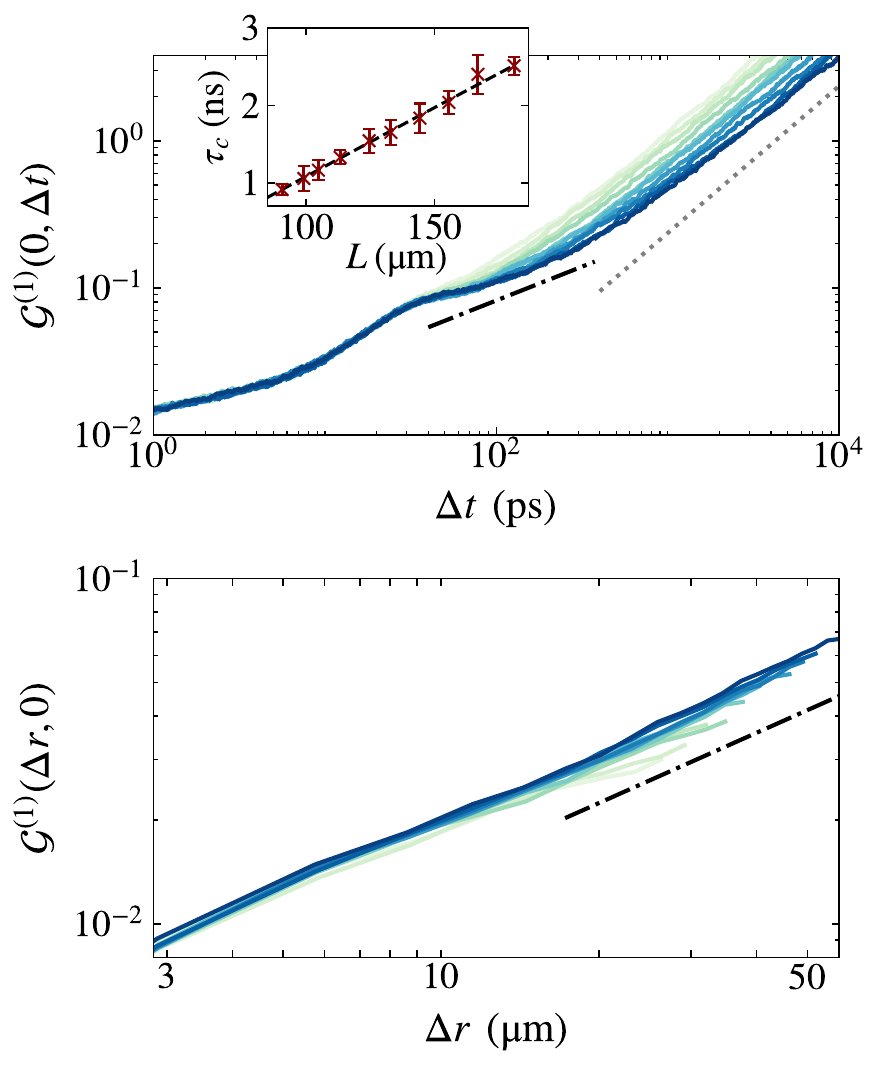}
\caption{\label{fig:finite_size_KPZ} First-order correlation function at the KPZ point of Fig.~\ref{fig:transition} for a system of $N=32$ to 64 linear sites, at (upper panel) equal-space, and (lower panel) equal-time. The colorscale is changing with increasing the system size from light green to dark blue. Stretched exponential decays are represented by the dash-dotted line. A grey dotted line indicates the large time exponential decay. Inset: coherence time $\uptau_c$ fitted from the large-time exponential decay of $g^{(1)}(0,\Delta t)$ as a function of the system size $L$.}
\end{figure}
When the nonlinearity $\lambda$ is large enough to let KPZ features arise, we expect $\tau_c\sim L^{0.83}$. Proceeding in the same manner as above, we compute the condensate first-order correlation function averaged over 2048 realizations at the KPZ point of Fig.~\ref{fig:transition}, varying N. As we have discussed in Sec.~\ref{sec:Method_phase_diag}, the position of the transition in Fig. \ref{fig:transition} depends on the size of the system, causing the phase turbulence regime where KPZ features are observed to disappear in favor of defect turbulence if the system size is too large. For this reason, N is varied from 32 to 64 only, including more intermediate points to improve the fit of the linewidth $\uptau_c(L)$. Results are shown on Fig.~\ref{fig:finite_size_KPZ}, where the spatial decoherence (lower panel) visibly converges to the KPZ scaling as the system size increases, as observed in Ref.~\cite{Mei_Wouters_scaling_2Dneq_EPBEC_2021}. The temporal decoherence (upper panel) shows that KPZ features are dramatically affected when the system gets smaller, the scaling window becoming hardly discernible. However, its signature remains accessible from the coherence time according to Eq.~\eqref{eq:scaling_finite_size}, whose dependence on the system size is close to linear (Inset of Fig.~\ref{fig:finite_size_KPZ}) and clearly differs from the parabola obtained on Fig.~\ref{fig:finite_size_EW}. Fitting the numerical values of the coherence time with a power law ansatz, we find  $\uptau_c\sim L^{0.97\pm0.18}$, which closely encloses the prediction from KPZ exponents and discard the one from EW exponents.

\subsection{\label{sec:Results_nk}Momentum distribution \texorpdfstring{$n_{\hat{\boldsymbol{k}}}$}{nk}}

\begin{figure}[ht!]
\includegraphics[width=7cm]{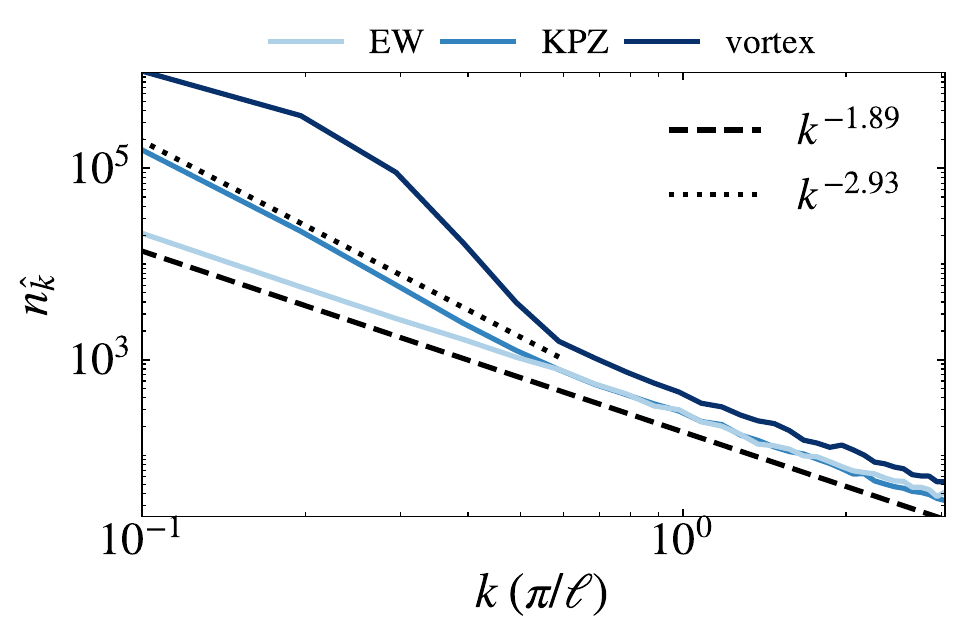}
\caption{\label{fig:nk}Momentum distributions corresponding to the EW, KPZ, and vortex points of Fig.~\ref{fig:transition} and averaged over 256 realizations of the noise. Dashed and dotted guidelines indicate the power laws $k^{-1.893\pm 0.004}$ and $k^{-2.93\pm0.03}$ respectively.}
\end{figure}

We now compute the momentum distribution of the condensate, defined as
\begin{equation}
    n_{\hat{\boldsymbol{k}}}=\langle \psi^*(\hat{\boldsymbol{k}}, t_0)\psi(\hat{\boldsymbol{k}}, t_0) \rangle,
    \label{eq:momentum_distrib}
\end{equation}
where $\psi(\hat{\boldsymbol{k}}, t_0)$ is the Fourier transform of the condensate wave function and $t_0$ is chosen in the stationary state. The results are displayed on Fig.~\ref{fig:nk} for the three points shown  on Fig.~\ref{fig:transition}. Let us first comment on the behavior of the momentum distribution for the large values of $k$. In the adiabatic approximation of the reservoir,  Bogoliubov theory predicts in the continuous limit that the momentum distribution is determined by the microscopic properties of the condensate according to $n_{\hat{\boldsymbol{k}}}\propto\sigma/\gamma_2 k^2$ (see Appendix~\ref{sec:ap_bogo}). We therefore expect the three curves to share the same decay at large $k$. This is indeed observed, and the momentum distribution is found to decay as $n_{\hat{\boldsymbol{k}}}\sim k^{-1.893\pm 0.004}$, represented on Fig.~\ref{fig:nk} by a dashed guideline, in good agreement with the theoretical prediction.

\medbreak
At small momenta, we expect $n_{\hat{\boldsymbol{k}}}$ to depend on the condensate macroscopic properties, so that it should behave differently in the three regimes under consideration. Within the conditions of validity of Bogoliubov theory, in particular for small density fluctuations, we find $n_{\hat{\boldsymbol{k}}}\propto D/\nu k^{2\beta z + 2}$. This leads to the prediction $n_{\hat{\boldsymbol{k}}}\propto D/\nu k^{2}$ at the EW point. On Fig.~\ref{fig:nk}, we observe {$n_{\hat{\boldsymbol{k}}}\sim k^{-1.893\pm 0.004}$ as in the large $k$ regime, which is in good agreement with the theoretical expectation. At the KPZ point, nonlinearity comes into play and the scaling is expected to change to $n_{\hat{\boldsymbol{k}}}\sim k^{-2.77}$. The observed behavior is fairly close to this prediction, as we obtain $n_{\hat{\boldsymbol{k}}}\sim k^{-2.93\pm 0.03}$ as shown by the dotted guideline.

In the vortex phase, density fluctuations are large and the momentum distribution is expected to display structures that depend on both the typical vortex radius and their average interdistance. The curve shown on Fig.~\ref{fig:nk} displays such a structure, which appears similar to the one observed in the soliton phase of 1d condensates \cite{vercesi2023phase}.

\subsection{\label{sec:Results_phase}Distribution of condensate phase and density fluctuations}

In this section, we compute the probability distributions of the fluctuations of the phase and of the density in the stationary state for the three  regimes highlighted on Fig.~\ref{fig:transition}. Let us start with the phase. As already mentioned in Sec.~\ref{sec:Univ_mappingKPZ}, interfaces in the KPZ universality class exhibit universal probability distributions of their fluctuations belonging to three main subclasses. This is encapsulated in the following asymptotic relation
\begin{equation}
    h(\boldsymbol{r_0}, t) \underset{t\to+\infty}{\sim} v_{\infty}t + (\Gamma t)^{\beta_{\rm KPZ}}\chi(\boldsymbol{r_0}, t),
    \label{eq:asymptotic_h}
\end{equation}
where the first term describes the deterministic growth of the interface, and where $v_{\infty}$ and $\Gamma$ are two non-universal constants whose exact expression depends on the underlying microscopic model \cite{Krugasymptoticvelocity}. The  second term in Eq.~\eqref{eq:asymptotic_h} encompasses the fluctuations, which grow in time with the KPZ scaling, and where $\chi$ is a random variable. Its probability distribution  depends on the geometry of the growing interface, distinguishing three universality subclasses that correspond to flat, curved or stationary geometries.  Each of them possesses a finite skewness and kurtosis, signaling a clear departure from the EW linear theory ($\lambda=0$) which is characterized by Gaussian fluctuations. These subclasses can be probed, in practice, through the distribution of reduced height fluctuations defined as $\Tilde{h}= \left[\Delta h(\Delta t+t_0) - \Delta h(t_0)\right]/(\Gamma \Delta t)^{\beta_{\rm KPZ}}$, with $\Delta h = h - \langle h \rangle$ at a fixed space point $\boldsymbol{r_0}$. For 1d interfaces, these probability distributions are given by Tracy-Widom GOE, Tracy-Widom GUE, and Baik-Rains distributions respectively \cite{corwin2012KPZ, takeuchi2018appetizer}. This was extensively confirmed for interfaces from both  numerical \cite{Takeuchi_JSM2012_numericsEden_GUE_Airy2_finite_time, Takeuchi_PRL2013_simuPNG_crossover_GOE_BR} and experimental studies \cite{Takeuchi_PRL2010_expe_GUE_Airy2, Takeuchi_JSP2012_expe_GOE_GUE_Airy1-2, Takeuchi_PRL2020_expe_BR_finite_time, Takeuchi_PRL2020_expe_flat-circular_crossover}. These three universality subclasses were also realized for the phase of 1d polariton condensates, adding an external confining potential  to induce a geometric constraint on the phase of the condensate \cite{squizzato2018KPZsubclasses, deligiannis2021KPZsubclasses}.

The same asymptotic form Eq.~\eqref{eq:asymptotic_h} holds in 2d, where three subclasses have also been identified in numerical simulations of models belonging to the 2d KPZ universality class. There are no known analytical forms for the corresponding distributions, and only empirical fitting functions are available. For the 2d analogs of the flat and curve interfaces, a generalized Gumbel distribution of parameter $m=6$ and $9.5$ respectively has been proposed \cite{Carrasco_NJP_2014_2Dsubclasses_Airy, Oliveira_subclasses2DKPZ_2013, HH_PRL2012_2D_universal_distrib}, and for the stationary geometry  a Pearson type IV distribution \cite{Halpin-Healy_PRE2013_Pearson_distribKPZ}. This seemed to be confirmed by experiments  although 2d growing interfaces are not fully controlled \cite{almeida2014, HH_Palasantzas_EPL2014_2DKPZ_thin_films, Almeida2017}, and two of the three subclasses have been reported in the phase of 2d EP BECs \cite{deligiannis2022KPZ2D, Ferrier_Marzena_2022}.

\medbreak
To confirm the scaling observed on Fig.~\eqref{fig:g1} at the KPZ point of Fig.~\eqref{fig:transition}, we record 5120 realizations of the condensate dynamics from which we extract phase trajectories in time in the steady state and unwrap them in $\mathbb{R}$ for a given spatial point. The presence of  defects creates time-resolved $2\pi$ jumps in the unwrapped phase, creating secondary peaks in the distribution of fluctuations \cite{FontaineNature2022, deligiannis2022KPZ2D}. However, very few phase jumps occur even for  large times for the parameters corresponding to the KPZ regime, so that the unwrapped phase can safely be considered as a continuous interface in the temporal window of interest. Typical trajectories are shown on Fig.~\ref{fig:typical_traj}.
\begin{figure}[ht!]
\includegraphics[width=7cm]{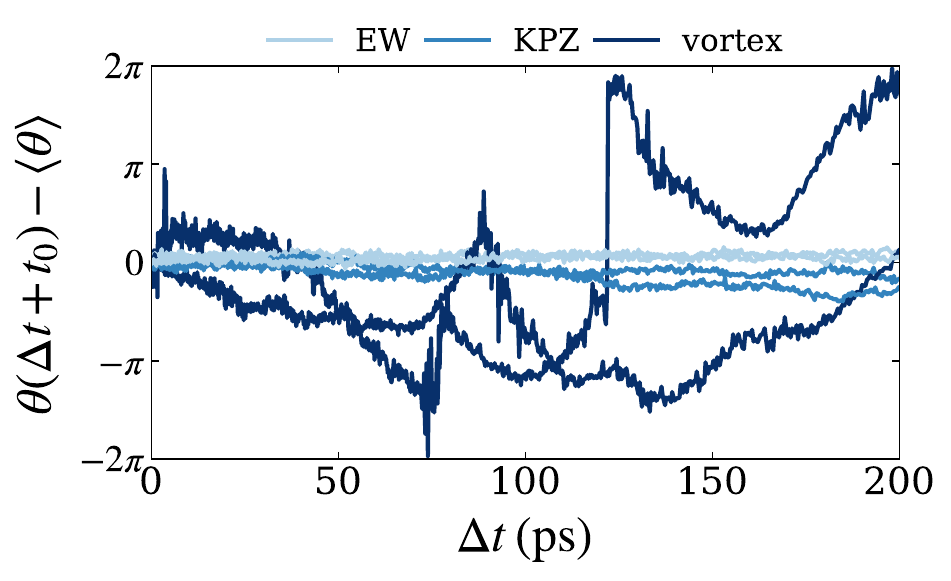}
\caption{\label{fig:typical_traj} Two typical unwrapped phase trajectories for each of the three regimes corresponding to the EW, KPZ, and vortex points of Fig.~\ref{fig:transition} (with the same color code). A $2\pi$ phase jump is visible on one of the vortex phase trajectories.}
\end{figure}
The normalized probability distribution of reduced phase fluctuations $\Tilde{\theta}$, defined by
\begin{equation}
    \Tilde{\theta}(\Delta t, t_0) = \text{sign}(\lambda)\dfrac{\Delta \theta(\Delta t + t_0) - \Delta \theta(t_0)}{\text{Var}\left[ \Delta \theta(\Delta t + t_0) - \Delta \theta(t_0) \right]}
    \label{eq:reduced_phase}
\end{equation}
where $\Delta t\in\left[35{\rm ps}, 350{\rm ps} \right]$ as shown on Fig.~\ref{fig:g1}, is displayed on Fig.~\ref{fig:distrib_phase}. The $\text{sign}(\lambda)$ in \eqref{eq:reduced_phase} stems for the negative value obtained for $\lambda$ with our parameters, which flips the distribution into its mirror image compared with the theoretical predictions obtained for $\lambda >0$ \cite{squizzato2018KPZsubclasses, FontaineNature2022}. The distribution of phase fluctuations exhibits clear non-Gaussian tails. As $\Delta t/t_0\sim0.1$ in our simulations, we expect it  to be described by the Pearson distribution. The latter is represented with a dashed blue line together with our numerical data on Fig.~\ref{fig:distrib_phase}, showing a perfect agreement with the expected KPZ distribution.

Proceeding in the same way for the EW point of Fig.~\ref{fig:transition} we obtain a Gaussian distribution instead (light blue histogram on Fig.~\ref{fig:distrib_phase}), clearly confirming the  slow crossover between an EW phase and the KPZ phase presented earlier. Typical unwrapped phase trajectories at the EW point are shown on Fig.~\ref{fig:typical_traj}. Let us note that this result is in agreement with the one  found  in numerical simulations of the deterministic 2d CGLE \cite{Chate_PhysicaD1996_phase_turbulence_deterministic_2DCGLE_finite_size}, where only the EW could be observed since in the deterministic case, tremendously large spatiotemporal scales are needed for KPZ features to arise \cite{Pandit_PRE2020_KPZ_deterministic_KS}.

\begin{figure}[ht!]
\includegraphics[width=7cm]{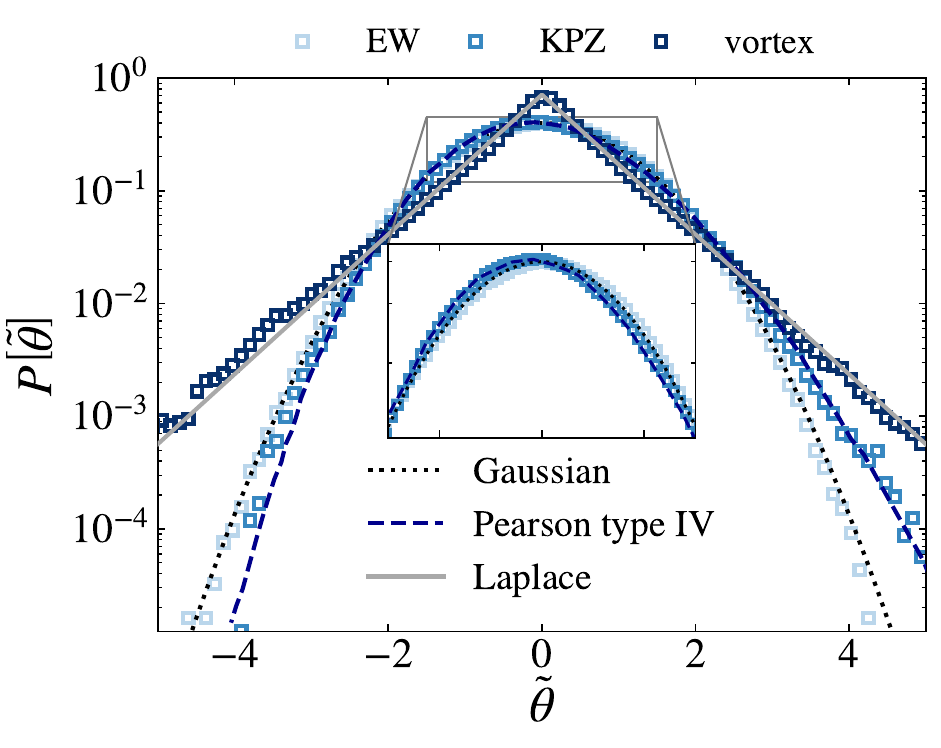}
\caption{\label{fig:distrib_phase} Centered and rescaled distributions of phase fluctuation $\Tilde{\theta}$ corresponding to the EW, KPZ, and vortex points of Fig.~\ref{fig:transition}. The two former histograms are composed of approximately $1.5\times 10^6$ data, and approximately $7\times10^6$ for the latter. The centered and unit-variance 2d stationary KPZ distribution \cite{Halpin-Healy_PRE2013_Pearson_distribKPZ} is shown by a dark blue dashed line, together with a centered and unit-variance Gaussian distribution. The grey plane line is a centered Laplace distribution of parameter $b=0.7$.}
\end{figure}

Following the same method for the vortex point of Fig.~\ref{fig:transition}, we compute the probability distribution of unwrapped phase fluctuations in the time window where the first-order correlation function is saturated. At this point, phase trajectories behave erratically and numerous defects are visible in Fig.~\ref{fig:typical_traj}. We obtain an histogram for $\Tilde{\theta}$ compatible with a double exponential distribution (dark blue squares on Fig.~\ref{fig:distrib_phase}). This indicates that phase jumps occur independently and at a constant rate in time, as was suggested in Refs.~\cite{Diehlspacetimevortex, vercesi2023phase}. This can be modeled by a Laplace distribution $\propto e^{-|\tilde\theta|/b}$, where we find $b\simeq 0.7$ from the fit of the numerical data.

\medbreak
We now turn to the study of density fluctuations. We extract for each of the three regimes shown on Fig.~\ref{fig:transition} the density fluctuations in the stationary regime from 128 independently produced spatial maps. Results are shown on Fig.~\ref{fig:distrib_density}, where we have normalized the density by its mean-field value $n_0$ for convenience.
\begin{figure}[ht!]
\includegraphics[width=7cm]{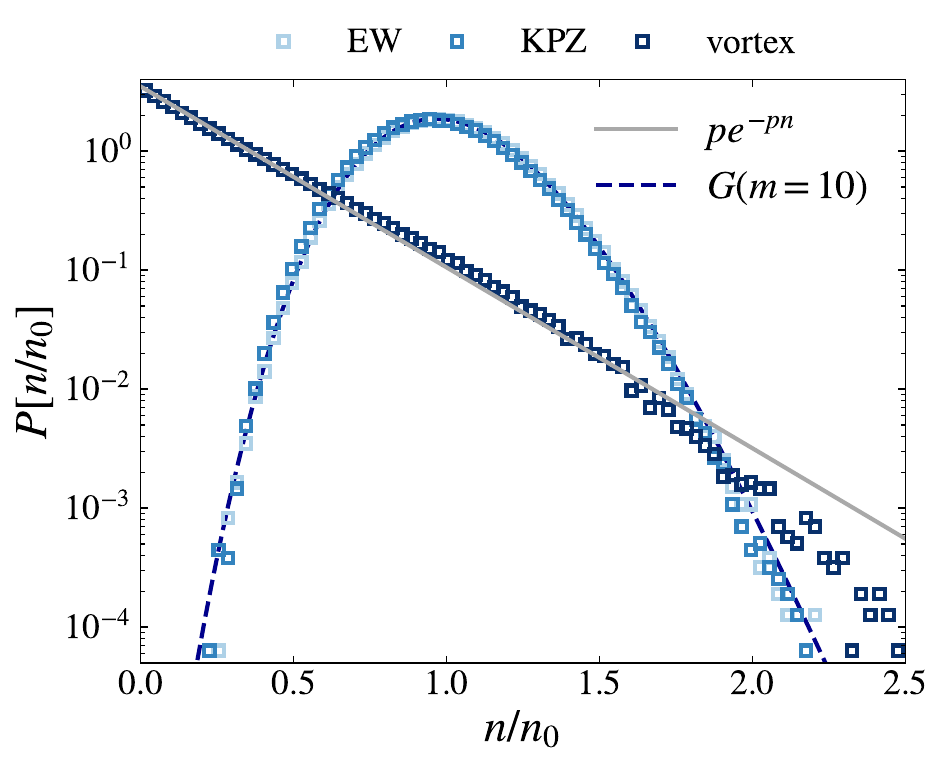}
\caption{\label{fig:distrib_density} Histograms of the density $n$ normalized by its mean-field value $n_0$ corresponding to the EW, KPZ, and vortex points of Fig.~\ref{fig:transition}. Each histogram contains approximately $5\times10^5$ data. The two formers are fitted by a Gumbel distribution of parameter $m=10$, mean $\langle n\rangle/n_0$, and variance $\text{Var}\left[ n/n_0 \right]$, shown by the dashed blue line. The latter is fitted by an exponential distribution of parameter $p=3.5$.}
\end{figure}
First we observe that the density distribution at the EW and at the KPZ point perfectly match, which is not too surprising from Figs.~\ref{fig:transition} and \ref{fig:g2}. 
{\color{black}
These distributions have asymmetric tails, stemming from the constraint $n\geq0$. In analogy with   photon lasers operating above threshold \cite{milburn-book}, one could expect the density fluctuations to have a Poisson distribution, but we find that our data  is not well described by such a distribution, and instead can be modeled by a Gumbel distribution of parameter $m=10$. Density fluctuations being related to that of the phase through \eqref{eq:gGPE}, it remains open whether or not KPZ fluctuations can leave a signature in the density. A definite conclusion requires further studies which are beyond the scope of this work.}
At the vortex point, the density distribution markedly differs and is rather fitted by an exponential distribution. This indicates again that the formation of defects occurs randomly, following a Poisson process. Similar exponential distributions have been reported in the nucleation time distribution of spiral defects, in the context of the 2d CGLE \cite{Liu_PRE2019_nucleation_defects_2DCGLE}, and the discrepancies on Fig.~\ref{fig:distrib_density} between the histogram and the exponential distribution at large $n/n_0$ can be explained by finite size effects \cite{Chate_PhysicaD1996_phase_turbulence_deterministic_2DCGLE_finite_size}.

\section{Conclusion}

In this paper, we have explored the universal properties of a 2d exciton-polariton condensate under incoherent pumping through numerical simulations of a generalized Gross-Pitaevskii equation. Controlling the effective nonlinearity of the phase dynamics through the interaction strength between reservoir excitons and polaritons, we have identified three different regimes: a first one for small nonlinearity where the dynamics of the density and of the phase are decoupled, and where the phase exhibits EW universal properties, a second one at intermediate nonlinearity where the density and phase are still decoupled but where the phase roughens with properties belonging to the KPZ universality class,  a third one for large values of the nonlinear parameter, where density and phase fluctuations are coupled, since spatial vortices induce  both phase singularities and deep minima in the condensate density at the positions of the vortex cores.

\medbreak
We have characterized the properties of these three regimes by studying the condensate first-order correlation function, the momentum distribution, and the distribution of the phase and density fluctuations.
In the weakly nonlinear limit,
we report  evidence of an algebraic decay of the first-order correlation function indicating the emergence of the EW universality class, from which we establish, upon increasing the interaction strength a slow crossover towards stretched exponential decay with exponents of the 2d KPZ class, and eventually a very rapid decay leading to saturation in the vortex phase. This observed behaviors in the EW and KPZ regimes are then confirmed by a careful
study of finite size effects in the condensate decoherence time.
The different phases  are also distinguished by their momentum density distribution, where  notably the EW and KPZ regimes exhibit different scaling at small momenta, in contrast with  1d condensates.

Finally, we have focused on the statistical properties of phase fluctuations to provide a further validation of the three identified regimes. We have found that the distribution of phase fluctuations is Gaussian in the EW
weakly nonlinear regime, it displays non-Gaussian tails at stronger non-linearity in the KPZ regime, in agreement  with those obtained in numerical simulations of interfaces in the 2d KPZ class, whereas the distribution is exponential in the vortex regime. Furthermore, we observe that the probability distribution of the density fluctuations {\color{black}does not have a simple form -- Gaussian or Poissonian -- in the EW and KPZ regimes.}

Our analysis highlights that earlier observations from experiments \cite{comaron_ostrovskaya_arXiv2024_algebraic_g1-r} and numerical simulations \cite{Comaron_Noneq_BKT_2021} correspond to the realization of the EW regime and provides a microscopic prediction for the observed power-law decays and the ratio of the corresponding exponents. {\color{black}This also clarifies the distinction between closed and dissipative condensates, while showing that quasi-order can be accessed in a weakly non-linear out-of-equilibrium configuration.}

Our work provides a unified picture of the universal properties of 2d polariton condensates in various parameter regimes. In outlook, it would be of great interest to provide a quantitative prediction for the transition to the vortex regime, whose qualitative picture has been developed in Refs.~\cite{Altman_cKPZ_duality, Altman_2016_EM_duality}, as well as a deeper understanding of oscillations observed in density correlations, and the extension of the proposed phase diagram to pumping rates close to the condensation threshold where large proliferation of space-time defects is expected.
It could be also interesting to further study the distributions of density fluctuations in the EW and KPZ regimes and provide a theoretical explanation of  the observed  statistics.
Overall, our findings pave the way for further experimental explorations of the universal properties of driven-dissipative quantum systems.

\begin{acknowledgments}
We warmly thank  Ivan Amelio, Marzena Szyma{\'n}ska, Harvey Weinberger and Vincent Rossetto for stimulating discussions.
FH acknowledges support from the Laboratoire d’excellence LANEF in Grenoble (ANR-10-LABX-51-01),  AM acknowledges funding from the Quantum-SOPHA ANR Project ANR-21-CE47-0009. LC acknowledges support  from Institut Universitaire de France (IUF). This work was partly supported  by the European Research Council (ERC) under the European Union's Horizon 2020 research and innovation programme (project ARQADIA, grant agreement no.~949730), and under Horizon Europe research and innovation programme (ANAPOLIS, grant agreement no.~101054448).
\end{acknowledgments}

\appendix

\section{\label{sec:ap_bogo}Adiabatic Bogoliubov theory }

\noindent
In this appendix, we explicit the connection between the phase of the condensate and the height of a growing interface, by relating its phase-phase correlation function to the EW two-point correlator under the Bogoliubov approximation. Adapting the calculations of Ref. \cite{Marzena2010_finite_size}, as detailed below, we show how the finite size of the system affects the long time decoherence of the EP BEC.

\noindent
Setting $\hbar = 1$, the condensate dynamical equations read
\begin{eqnarray}
i\partial_t\psi & = & \Bigg[ \mathcal{F}^{-1}\left[ \epsilon_{\hat{\boldsymbol{k}}} - \dfrac{i}{2}\gamma_{\ell}(\hat{\boldsymbol{k}}) \right] + \dfrac{i R}{2}n_R \nonumber\\
    & & +  g|\psi|^2 + 2 g_Rn_R  \Bigg]\psi + \xi \label{eq:ap_gGPE} \\
\partial_tn_R & = & P - \left(\gamma_R + R|\psi|^2 \right)n_R \label{eq:ap_xreservoir}\,,
\end{eqnarray}
where $\langle \xi \rangle=0$ and $\langle \xi(\boldsymbol{r},t)\xi^*(\boldsymbol{r'},t') \rangle=2\sigma\delta(\boldsymbol{r}-\boldsymbol{r'})\delta(t-t')$.

In the adiabatic approximation of the reservoir $\partial_tn_R\approx 0$, Eq.~\eqref{eq:ap_xreservoir} reduces to $n_R = P/(\gamma_R+R|\psi|^2)$. The condensate wavefunction is expanded in fluctuations $\delta\psi$ around its stationary mean field  density $n_0= \gamma_R (p-1)/R$, as $\psi(\boldsymbol{r},t) = e^{-i{\color{black}\mu_{\rm th}}t}\left[ \sqrt{n_0} + \delta\psi(\boldsymbol{r},t)\right]$. Injecting it in Eq.~\eqref{eq:ap_gGPE} leads at zeroth order to the mean field exciton reservoir density and condensate blueshift, respectively $n_{R0}=\gamma_0/R$ and ${\color{black}\mu_{\rm th}} = gn_0 + 2g_Rn_{R0}$. To first order in $\delta\psi$, Eq.~\eqref{eq:ap_gGPE} reduces to the linear dynamics
\begin{eqnarray}
    i\partial_t\begin{pmatrix}
    \delta\psi \\
    \delta\psi^*
    \end{pmatrix} = \mathcal{F}^{-1}\left[\mathcal{L}\right]\begin{pmatrix}
    \delta\psi \\
    \delta\psi^*
    \end{pmatrix} + \sqrt{\sigma}\begin{pmatrix}
\Tilde{\xi} \\
-\Tilde{\xi}
\end{pmatrix}\,
\label{eq:ap_Bogo_psi}
\end{eqnarray}
where we introduced $\Tilde{\xi} = \xi e^{-i{\color{black}\mu_{\rm th}}t}$ and the Bogoliubov matrix
\begin{eqnarray}
    \mathcal{L}=
    \begin{pmatrix}
        \epsilon_{\hat{\boldsymbol{k}}} + g_{e}n_0 + i\Gamma_{\hat{\boldsymbol{k}}} & g_{e}n_0 - ig_{i}n_0 \\
        - g_{e}n_0 - ig_{i}n_0 & -\left( \epsilon_{\hat{\boldsymbol{k}}} + g_{e}n_0  \right) - i\Gamma_{\hat{\boldsymbol{k}}}
    \end{pmatrix}\,
    \label{eq:ap_Bogomatrix_L}
\end{eqnarray}
with $\Gamma_{\hat{\boldsymbol{k}}} = g_{i}n_0 + \frac{\Delta\gamma(\hat{\boldsymbol{k}})}{2} $, $\Delta\gamma(\hat{\boldsymbol{k}}) = \gamma_{\ell}(\hat{\boldsymbol{k}}) - Rn_{R0}$ and $g_{i}=\gamma_0^2/2P$. 

\medbreak
We introduce the rotation matrix 
\begin{equation}
    \mathcal{R} = \dfrac{1}{2\sqrt{n_0}}
    \begin{pmatrix}
    1 & 1 \\
    -i & i
    \end{pmatrix}
    \label{eq:ap_A}
\end{equation} 
such that
\begin{eqnarray}
    \begin{pmatrix}
    \delta n/2n_0 \\
    \delta\theta
    \end{pmatrix} = \mathcal{R}
    \begin{pmatrix}
    \delta\psi \\
    \delta\psi^*
    \end{pmatrix}
\label{eq:ap_transfo_phi_n_theta}
\end{eqnarray}
and 
\begin{eqnarray}
    \left(\partial_t - \mathcal{F}^{-1}\left[\mathcal{L}_{rot}\right]\right)\begin{pmatrix}
    \delta n/2n_0 \\
    \delta\theta
    \end{pmatrix} =\sqrt{\dfrac{\sigma}{n_0}}\begin{pmatrix}
-{\rm Im}\left(\Tilde{\xi}\right) \\
{\rm Re}\left(\Tilde{\xi}\right)
\end{pmatrix}\, ,
\label{eq:ap_Bogo_n_theta}
\end{eqnarray}
with
\begin{equation}
    \mathcal{L}_{rot} = \begin{pmatrix}
        -g_{i}n_0 - \Gamma_{\hat{\boldsymbol{k}}}   &  \epsilon_{\hat{\boldsymbol{k}}}  \\
        -\left( \epsilon_{\hat{\boldsymbol{k}}} + 2g_{e}n_0 \right) & g_{i}n_0 - \Gamma_{\hat{\boldsymbol{k}}} 
    \end{pmatrix}\, .
    \label{eq:ap_Lrot}
\end{equation}
The eigenvalues of this matrix display the characteristic double-branched spectrum 
\begin{equation}
    \omega_{\pm} = -i\Gamma_{\hat{\boldsymbol{k}}} \pm \sqrt{E_{\hat{\boldsymbol{k}}}^2 - g_{i}^2n_0^2}\, ,
    \label{ap:bogoEigenval}
\end{equation}
with $E_{\hat{\boldsymbol{k}}}^2 = \epsilon_{\hat{\boldsymbol{k}}}(\epsilon_{\hat{\boldsymbol{k}}} + 2g_{e}n_0)$.

\subsection{Phase-phase correlator}

Within this approximation, the phase correlation reads in Fourier space
\begin{eqnarray}
    \langle \delta\theta_{\boldsymbol{k},\omega}\delta\theta_{-\boldsymbol{k},-\omega} \rangle & = & \dfrac{4\sigma}{n_0}\dfrac{g_en_0\left(\epsilon_{\hat{\boldsymbol{k}}} + g_en_0\right) - g_in_0\Gamma_{\hat{\boldsymbol{k}}}}{(\omega^2 - \omega_-^2)(\omega^2 - \omega_+^2)} \nonumber \\
    &  & + \dfrac{\langle \delta n_{\boldsymbol{k},\omega}\delta n_{-\boldsymbol{k},-\omega} \rangle}{4n_0^2} \label{ap:eq_thetatheta_k_w_general}, \\
    {\color{black}\langle \delta n_{\boldsymbol{k},\omega}\delta n_{-\boldsymbol{k},-\omega} \rangle} & = & {\color{black}\dfrac{\sigma n_0(\epsilon_{\boldsymbol{\hat{k}}}^2 + \omega^2 + (\Gamma_{\boldsymbol{\hat{k}}} + g_i n_0)^2)}{(\omega^2 - \omega_-^2)(\omega^2 - \omega_+^2)} \label{ap:eq_nn_k_w_general}.}\nonumber
\end{eqnarray}

The large scale properties of the condensate can be understood taking advantage of the splitting between these two branches at the crossover momentum $k_c=\sqrt{2n_0|m|}\left[ \sqrt{g_e^2 + g_i^2} - |g_e| \right]^{1/2}$, separating the infrared collective modes $k\ll k_c$ from the ultraviolet free particles $k\gg k_c$. Focusing on the former limit and choosing $\epsilon_{\hat{\boldsymbol{k}}}=k^2/2m$ and $\gamma_{\ell}(\hat{\boldsymbol{k}}) = \gamma_0 + \gamma_2 k^2$, we find for $k\approx0$ that $\omega_{-}\approx -2 i g_in_0$ and $\omega_{+}\approx -i\nu k^2$, where $\nu$ is the effective phase viscosity \eqref{eq:paramKPZ_nu} introduced in the main text.

Under the condition that $k\ll\sqrt{2g_in_0/\nu}$ (for small $\mu_{\rm th}$, $\sqrt{2g_in_0/\nu} \sim 10 k_c$), the real space two-point correlation function $C_{\theta\theta}(r, t) = \langle \left[ \delta\theta(\boldsymbol{r}, t) - \delta\theta(\boldsymbol{0}, 0) \right]^2 \rangle$ reads in generic dimension $d$ 
\begin{eqnarray}
    C_{\theta\theta}(r, t) & \approx & C_{\rm EW}(r,t) - \dfrac{C_{nn}(r,t)}{2n_0^2} \nonumber \\ 
    && - 2D\int \dfrac{d\boldsymbol{k}}{(2\pi)^d}\dfrac{1-e^{i\boldsymbol{k}\cdot\boldsymbol{r}-2g_in_0t}}{2g_in_0} \nonumber \\
    && + \dfrac{8\sigma g_in_0\nu}{n_0}\int\dfrac{d\boldsymbol{k}d\omega}{(2\pi)^{d+1}} (1-e^{i(\omega t + \boldsymbol{k}\cdot\boldsymbol{r})}) \nonumber \\
    && \times \dfrac{k^2}{\left(\omega^2 + 4g_i^2n_0^2\right)\left( \omega^2 + \nu^2 k^4 \right)}
    \label{ap:eq_Cthetatheta_EW_nn}
\end{eqnarray}
where $C_{nn}(r,t) = \langle \delta n (r,t) \delta n (0,0)\rangle - \langle \delta n (0,0)^2 \rangle$. The first term, which is the Edwards-Wilkinson two-point correlation defined by \eqref{eq:ap_EW_correlation}, 
dominates the scaling behavior of $C_{\theta\theta}(r,t)$ {\color{black}in the infrared regime. }
{\color{black} This can be established by integrating over the frequency, and  analyzing the small $k$ behavior of the integrals at fixed $t$ and  $r$. One finds that, for all the terms but $C_{\rm EW}(r,t)$, the integrals are convergent in the IR, hence they vanish at large $r$. They are dominated by the UV cut-off $\Lambda$. In contrast, in the EW contribution, the leading term at small $k$ diverges in the IR as $\sim \int dk/k$ for $d=2$ and thus dominates the IR behavior, as is detailed below.}

  The EW correlation reads \cite{edwards1982surface, Nattermann_PRA1992_solutionEW}
\begin{equation}
    C_{\rm EW}(r,t) = \mathcal{C}_0 t^{2\beta} F^{(d)}\left( \dfrac{r}{y_0t^{1/z}} \right) 
    \label{eq:ap_EW_correlation}
\end{equation}
where $\mathcal{C}_0 = 2\frac{D}{\nu}(4\nu)^{2\beta}$ and $y_0 = 2\sqrt{\nu}$ are non-universal constants, with $D$ and $\nu$ the effective parameters defined in Eqs.~\eqref{eq:paramKPZ_nu} and \eqref{eq:paramKPZ_D}. This function is characterized by the universal dynamical exponent $z=2$, the growth exponent $\beta = \frac{2-d}{4}$, and the universal scaling function \cite{barabasi1995fractal, Nattermann_PRA1992_solutionEW}.
{\color{black}
\begin{equation}
F^{(d)}(y) = \int \frac{d\boldsymbol{\tilde{k}}}{(2\pi)^d} \frac{1-e^{i\boldsymbol{\tilde{k}}\cdot\boldsymbol{y} - \boldsymbol{\tilde{k}}^2/4}}{\boldsymbol{\tilde{k}}^2}\quad \hbox{with}\quad \boldsymbol{\tilde{k}} = 2\sqrt{\nu t}\boldsymbol{k}\end{equation}}

\medbreak
Let us analyze the finite-size effects  on $C_{\rm EW}(r=0,t)$. For a system  of linear size $L$, from the diffusive nature of EW modes, one
expects a change in its behavior at a time $t^*\sim L^2/\nu$. Adapting the arguments given in Ref.~\cite{Marzena2010_finite_size}, at fixed $t$, one
can split the dimensionless momentum integral of $F^{(d)}(y)$ into two: one for ${k}\leq 1/\sqrt{\nu t}$, the other for $k> 1/\sqrt{\nu t}$.

{\color{black}
In the limit of small times $t\ll t^*$, the bound  $1/\sqrt{\nu t}$ is large, such that both integrals  contain enough modes to be treated as continuous integrals. In this case,
 in dimension $d = 1$,
 the integrals can be performed analytically, yielding $C_{\rm EW}(0, t\ll t^*) = \dfrac{2D}{\sqrt{\pi\nu}}\sqrt{t}$ consistent with the scale invariant form \eqref{ap:eq_Cthetatheta_EW_nn}.
In dimension $d = 2$, one can expand to second order  $e^{-\tilde{k}^2/4}$ in the first integral and send the UV cutoff $\Lambda$ to infinity in the convergent part $\propto e^{-\tilde{k}^2/4}$ of the second integral. Both contributions are finite, and the result is thus dominated in the IR by the  $\sim \int dk/k$  contribution, leading to  $C_{\rm EW}(0, t) \approx \dfrac{D}{2\pi\nu}\ln(\nu t \Lambda^2)$, {\it i.e.} a logarithmic time dependence \cite{Forrest1990, PAL1999_EW2+1, Bray_1994_phase_ordering_kinetics, Nattermann_PRA1992_solutionEW, Tang_PRA1992_hybercube_stacking_EW-KPZ_crossover} with $\Lambda$ a momentum cutoff.

In the limit of large times $t\gg t^*$, the bound $1/\sqrt{\nu t}$ tends to zero. In this case, the first integral should be replaced by the discrete sum over the lowest modes. Expanding $e^{-\tilde{k}^2/4}$, one obtains that in both $d=1$ and 2, this term yields the dominant contribution at large time which is proportional to $t$. In $d=1$, it reads $C_{\rm EW}(0, t) \approx \dfrac{2D}{\mathcal{L}}t$ where $\mathcal{L}$ is a length scale maintaining homogeneity. A natural choice for this length scale, stemming from the lowest momenta of the finite-size system, is $\mathcal{L} = \uptau^{(1)}L$, with $\uptau^{(1)}$ a dimensionless constant \cite{amelio2024KPZlinewidth}. This constant contains the dependence of the condensate linewidth on the microscopic parameters that goes beyond the Bogoliubov approximation. In $d=2$, one obtains
 $C_{\rm EW}(0, t) \approx \dfrac{D}{\uptau^{(2)}L^2}t$ with the dimensionless constant $\uptau^{(2)}$.}

\medbreak
Overall, these results derived from the microscopic model reveal that Edwards-Wilkinson universal properties fully explain the decoherence of the condensate reported in the main text for small $\mu_{\rm th}$. They can be recasted under the following scaling form
\begin{equation}
    C_{\theta\theta}(0,t) \sim  t^{2\beta}f^{(d)}\left( \dfrac{t}{L^z} \right).
    \label{ap:eq_scaling_finite_size}
\end{equation}
with $f^{(d)}(y)\underset{y\to+\infty}{\sim}y^{1-2\beta}$ for any dimensions, while the limit $f^{(d)}(y\to0)$ depends on it as suggested in Ref.~\cite{Pandit_PhysicaD1996_conjeture_PT_CGLE_finite_size}. The scaling form \eqref{ap:eq_scaling_finite_size} was found to also hold when $\lambda\neq0$, in numerical simulations of 1d systems, where the exponents $z$ and $\beta$ are those of the KPZ class \cite{Krug_AP1997_finite_size, amelio2024KPZlinewidth}.

From this finite-size study, we expect $g^{(1)}(t\gg t^*)\sim e^{-t/\uptau_c(L)}$ with the coherence time $\uptau_c(L)\propto L^{z-2\beta z}$, generalizing the result of \cite{amelio2022Bogo_linewidth, amelio2024KPZlinewidth} to 2d systems. This Schawlow-Townes-like limit \cite{Schawlow_1958}, following EW or KPZ features, means that the growing phase interface looses the memory of its initial condition at long times and behaves as a Brownian motion in time \cite{Fabre_2010}.

\subsection{Momentum distribution}

From Eq.~\eqref{eq:ap_Bogo_psi}  after integrating over frequencies, we immediately get the momentum distribution $n_{\hat{\boldsymbol{k}}}=\langle \psi^*(\hat{\boldsymbol{k}}, t)\psi(\hat{\boldsymbol{k}}, t) \rangle$ as
\begin{equation}
    n_{\hat{\boldsymbol{k}}} = (2\pi)^d n_0 \delta(\boldsymbol{k}) + \dfrac{\sigma}{\Gamma_{\hat{\boldsymbol{k}}}}\left[ 1 + n_0^2\dfrac{g_e^2 + g_i^2}{E_{\hat{\boldsymbol{k}}}^2 + \Gamma_{\hat{\boldsymbol{k}}}^2 - g_i^2n_0^2} \right].
    \label{eq:ap_nk}
\end{equation}
Replacing $\epsilon_{\hat{\boldsymbol{k}}}=k^2/2m$ and $\gamma_{\ell}(\hat{\boldsymbol{k}}) = \gamma_0 + \gamma_2 k^2$, we obtain $n_{\hat{\boldsymbol{k}}}\underset{k\to+\infty}{\sim}\sigma/\gamma_2 k^2$ in the ultraviolet regime. The short-momentum behavior is given, up to a constant, by $n_{\hat{\boldsymbol{k}}}\underset{k\to 0^+}{\sim} \Tilde{C}_{\rm EW}(k, t=0)$, where $\Tilde{C}_{\rm EW}(k,t) = \int d\boldsymbol{r}e^{-i\boldsymbol{k}\cdot\boldsymbol{r}}C_{\rm EW}(r,t)$.

At equal time, we obtain $\Tilde{C}_{\rm EW}(k,0)\sim k^{-2\beta z - d}$. From this, we deduce that in the EW regime $n_{\hat{\boldsymbol{k}}}\propto k^{-2}$. {\color{black} This tail in the momentum distribution is similar to that of a thermal closed Bose gas, from which an effective polariton temperature can be defined as discussed in the main text.}
Since this result was obtained using the linear theory, we can expect the non-linearity of the phase dynamics to affect this regime and rather lead to $n_{\hat{\boldsymbol{k}}}\underset{k\to 0^+}{\sim} \Tilde{C}_{\rm KPZ}(k, t=0)\sim k^{-2\beta z - d}$ with the KPZ exponents instead, giving $n_{\hat{\boldsymbol{k}}}\propto k^{-2.77}$. This naive guess coincides with the KPZ asymptotic behavior observed in simulations of the noisy Kuramoto-Sivashinsky equation \cite{Cuerno_KPZasympotics_2DKS}, as well as with predictions from acoustic turbulence in a 2d superfluid Bose gas \cite{Pawlowski_PRA2015_nk_driven_BEC_2-3D}. {\color{black}This departure from the thermal-like tail $n_{\hat{\boldsymbol{k}}}\propto k^{-2}$ can be interpreted as a signature of the purely non-equilibrium nature of the KPZ equation in 2d.}

\medbreak
Note that for $d=1$, both EW and KPZ universality classes produce the same low-momentum power-law decay $n_{\hat{k}}\propto 1/k^2$, as confirmed by numerical simulations \cite{vercesi2023phase}
{\color{black}. This is a consequence of the accidental time-reversal symmetry of the KPZ equation in $d=1$, which yields a fluctuation-dissipation relation and leads to the same equal-time behavior as the EW equation, although the temporal beahvior is distinct. This time-reversal symmetry no longer holds in $d\neq 1$ for the KPZ equation \cite{Canet2010}.}



\bibliographystyle{prsty}
\providecommand{\noopsort}[1]{}\providecommand{\singleletter}[1]{#1}%
%

\end{document}